\documentclass[prb,twocolumn,showpacs,amsmath,amssymb,superscriptaddress]{revtex4-2}
\usepackage{dcolumn}% Align table columns on decimal point
\usepackage{bm,graphicx}% bold math
\usepackage{etoolbox}
\usepackage{url}
\usepackage{booktabs}
\usepackage[colorlinks]{hyperref}
\usepackage{breakurl}
\usepackage{color}
\usepackage [latin1]{inputenc}
\usepackage{siunitx}

\newsavebox\myboxA
\newsavebox\myboxB
\newlength\mylenA

\newcommand*\xoverline[2][0.75]{%
    \sbox{\myboxA}{$\m@th#2$}%
    \setbox\myboxB\null% Phantom box
    \ht\myboxB=\ht\myboxA%
    \dp\myboxB=\dp\myboxA%
    \wd\myboxB=#1\wd\myboxA% Scale phantom
    \sbox\myboxB{$\m@th\overline{\copy\myboxB}$}%  Overlined phantom
    \setlength\mylenA{\the\wd\myboxA}%   calc width diff
    \addtolength\mylenA{-\the\wd\myboxB}%
    \ifdim\wd\myboxB<\wd\myboxA%
       \rlap{\hskip 0.5\mylenA\usebox\myboxB}{\usebox\myboxA}%
    \else
        \hskip -0.5\mylenA\rlap{\usebox\myboxA}{\hskip 0.5\mylenA\usebox\myboxB}%
    \fi}
\makeatother
\usepackage[export]{adjustbox}
\usepackage{float}
\usepackage{siunitx}
\usepackage{booktabs}
\usepackage{color}
\usepackage{xcolor}
\usepackage [latin1]{inputenc}
\usepackage{dcolumn}% Align table columns on decimal point
\usepackage{etoolbox}
\usepackage{url}
\usepackage[colorlinks]{hyperref}
\usepackage[normalem]{ulem}
\usepackage{array}
\newcolumntype{x}[1]{>{\centering\arraybackslash\hspace{0pt}}p{#1}}
\usepackage{multirow}
\usepackage{accents}
\usepackage{braket}

\usepackage[normalem]{ulem}

\begin{document}

\title{Electronic band structure of Sb$_2$Te$_3$}

\author{I.~Mohelsky}
\affiliation{Laboratoire National des Champs Magn\'etiques Intenses, LNCMI-EMFL, CNRS UPR3228, Univ.~Grenoble Alpes, Univ.~Toulouse, Univ.~Toulouse~3, INSA-T, Grenoble and Toulouse, France}

\author{J.~Wyzula}
\affiliation{Laboratoire National des Champs Magn\'etiques Intenses, LNCMI-EMFL, CNRS UPR3228, Univ.~Grenoble Alpes, Univ.~Toulouse, Univ.~Toulouse~3, INSA-T, Grenoble and Toulouse, France}
\affiliation{Scientific Computing, Theory and Data Division, Paul Scherrer Institute, Switzerland}

\author{F.~Le Mardel\'e}
\affiliation{Laboratoire National des Champs Magn\'etiques Intenses, LNCMI-EMFL, CNRS UPR3228, Univ.~Grenoble Alpes, Univ.~Toulouse, Univ.~Toulouse~3, INSA-T, Grenoble and Toulouse, France}

\author{F. Abadizaman}
\affiliation{Department of Condensed Matter Physics and Central European Institute of Technology, Masaryk University, K\!otl\'a\v rsk\'a 2, 611 37 Brno, Czech Republic}

\author{O.~Caha}
\affiliation{Department of Condensed Matter Physics and Central European Institute of Technology, Masaryk University, K\!otl\'a\v rsk\'a 2, 611 37 Brno, Czech Republic}

\author{A.~Dubroka}
\affiliation{Department of Condensed Matter Physics and Central European Institute of Technology, Masaryk University, K\!otl\'a\v rsk\'a 2, 611 37 Brno, Czech Republic}

\author{X.~D.~Sun}
\affiliation{Laboratoire National des Champs Magn\'etiques Intenses, LNCMI-EMFL, CNRS UPR3228, Univ.~Grenoble Alpes, Univ.~Toulouse, Univ.~Toulouse~3, INSA-T, Grenoble and Toulouse, France}

\author{C.~W.~Cho}
\affiliation{Laboratoire National des Champs Magn\'etiques Intenses, LNCMI-EMFL, CNRS UPR3228, Univ.~Grenoble Alpes, Univ.~Toulouse, Univ.~Toulouse~3, INSA-T, Grenoble and Toulouse, France}

\author{B.~A.~Piot}
\affiliation{Laboratoire National des Champs Magn\'etiques Intenses, LNCMI-EMFL, CNRS UPR3228, Univ.~Grenoble Alpes, Univ.~Toulouse, Univ.~Toulouse~3, INSA-T, Grenoble and Toulouse, France}

\author{M.~F.~Tanzim}
\affiliation{Institute for Theoretical Physics Amsterdam, University of Amsterdam, and European Theoretical Spectroscopy Facility (ETSF), 1090 GL Amsterdam, The Netherlands}

\author{I.~Aguilera}
\affiliation{Institute for Theoretical Physics Amsterdam, University of Amsterdam, and European Theoretical Spectroscopy Facility (ETSF), 1090 GL Amsterdam, The Netherlands}

\author{G.~Bauer}
\affiliation{Institut f\"ur Halbleiter- und Festk\"orperphysik, Johannes Kepler Universit\"at, Altenbergerstrasse 69, 4040 Linz, Austria}

\author{G.~Springholz}
\affiliation{Institut f\"ur Halbleiter- und Festk\"orperphysik, Johannes Kepler Universit\"at, Altenbergerstrasse 69, 4040 Linz, Austria}

\author{M.~Orlita}
\email{milan.orlita@lncmi.cnrs.fr}
\affiliation{Laboratoire National des Champs Magn\'etiques Intenses, LNCMI-EMFL, CNRS UPR3228, Univ.~Grenoble Alpes, Univ.~Toulouse, Univ.~Toulouse~3, INSA-T, Grenoble and Toulouse, France}
\affiliation{Faculty of Mathematics and Physics, Charles University, Ke Karlovu 5, Prague, 121 16, Czech Republic}
    
\date{\today}

\begin{abstract} 
Here we report on Landau level spectroscopy of an epitaxially grown thin film of the topological insulator Sb$_2$Te$_3$, complemented by ellipsometry and magneto-transport measurements. The observed response suggests that Sb$_2$Te$_3$ is a direct-gap semiconductor with the fundamental band gap located at the $\Gamma$ point, or along the trigonal axis, and its width reaches $E_g = (190 \pm 10)$~meV at low temperatures. Our data also indicate the presence of other low-energy extrema with a higher multiplicity in both the conduction and valence bands.  The conclusions based on our experimental data are confronted with and to a great extent corroborated by the electronic band structure calculated using the $GW$ method. 
\end{abstract}

\maketitle

\section{Introduction}

The family of materials based on pnictogen and chalcogen atoms comprises a number of appealing topological systems~\cite{HasanRMP10,QiRMP11}. Among others, it also includes narrow-gap semiconductors Bi$_2$Se$_3$, Bi$_2$Te$_3$ and Sb$_2$Te$_3$ which were proposed~\cite{ZhangNaturePhys09} and confirmed as three-dimensional  topological insulators at very early a stage of the topological epoch~\cite{HsiehNature09,ChenScience09,HsiehPRL09}. Nowadays they belong to the best-known examples of systems with a non-trivial topology of the electronic band structure. 

Lately emerged interest in topological aspects of matter, together with a huge progress in experimentation of surface-sensitive techniques (ARPES and STM/STS, in particular) caused, that our understanding of bulk electronic band structure of these materials often lags behind their intriguing surface states. Only recent results of Landau-level (LL) spectroscopy measurements~\cite{OrlitaPRL15,MohelskyPRB20}, combined with data from early magneto-transport experiments~\cite{Kohlerpssb76,Kohlerpssb76II,Kohlerpssb76III}, clarified that Bi$_2$Se$_3$ and Bi$_2$Te$_3$ are both direct-gap semiconductors, the latter with a multiple valley degeneracy ($N=6$). 

\begin{figure}
      \includegraphics[width=.43\textwidth]{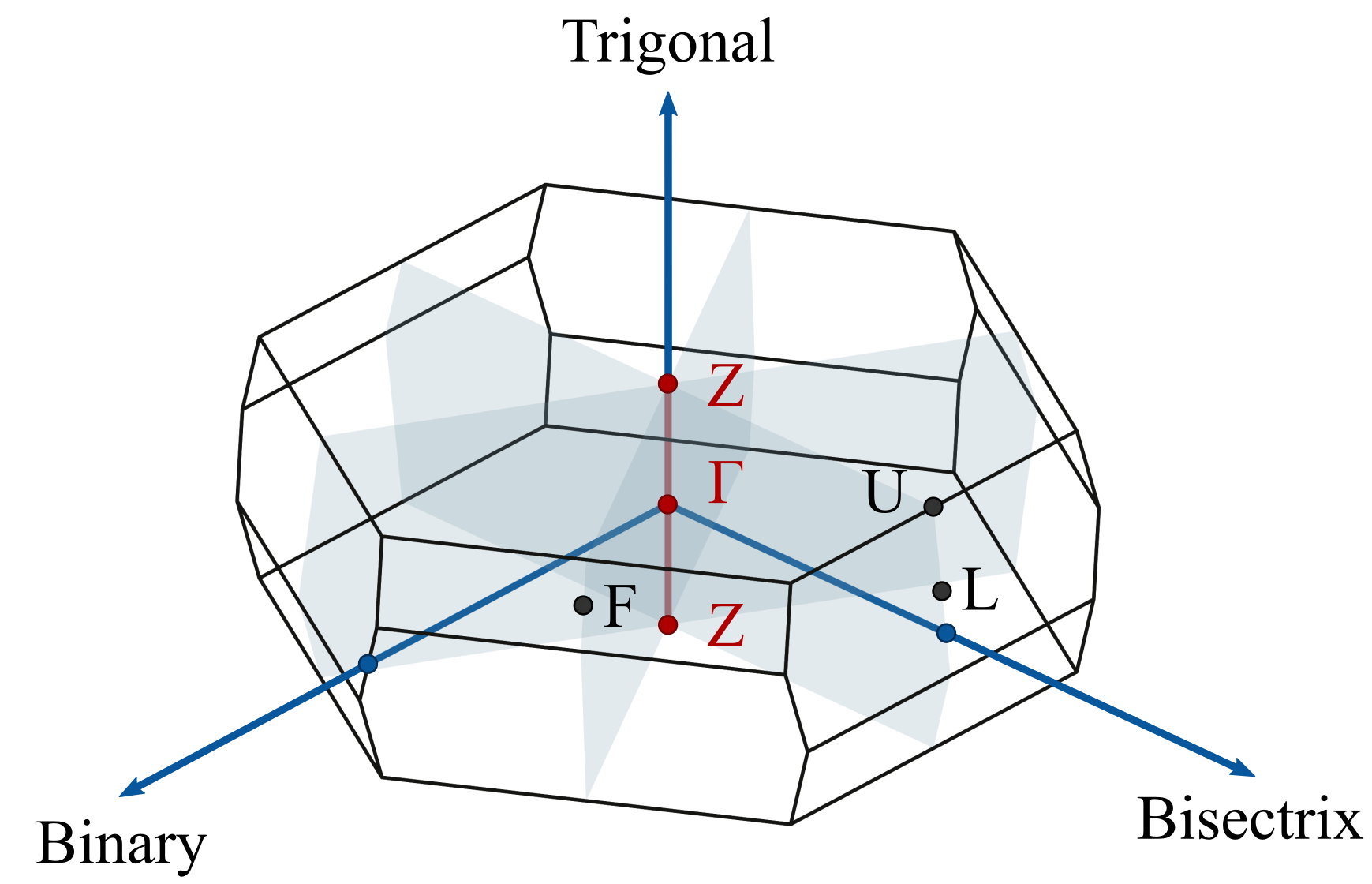}
      \caption{\label{BZ} Schematic view of the first BZ of Sb$_2$Te$_3$ with the mirror planes (gray planes) and trigonal, bisectrix as well as binary axes indicated.}
\end{figure}

As compared to Bi$_2$Se$_3$ and Bi$_2$Te$_3$, the bulk band structure of Sb$_2$Te$_3$ remains to a great extent uncovered. The quantum oscillations experiments have so far been performed only on bulk $p$-type samples with rather high hole concentrations (above $10^{20}$~cm$^{-3}$). The interpretation suggested six-fold~\cite{SchwartzSSC67,Kohlerpss77,KulbachinskiiPRB95,ZhaoPRB19,KulbachinskiiMTP21} or twelve-fold~\cite{MiddendorffSSC73} valley degeneracy, but also a more complex structure of the valence band, with six- and non-degenerated valleys combined~\cite{Simonpssb81}. While the single-valley degeneracy is connected with the center of the Brillouin zone (BZ) only, the sixfold degenerated band extrema may be located along the binary axes,
or in the mirror planes, see Fig.~\ref{BZ}. 

The band structure of Sb$_2$Te$_3$ has been also several times addressed theoretically~\cite{WangPRB07, ZhangNaturePhys09, YavorskyPRB11, AguileraPRB13, NechaevPRB15, AguileraPRB19}. The results of various ab-initio approaches are not fully consistent, but at least indicate that the band extrema should be primarily located at the $\Gamma$ point, along the $Z-\Gamma-Z$ line or in the mirror planes. Notably, the calculated energies of different conduction or valence band extrema often differ only by several tens of meV. This is likely beyond the precision of theoretical methods and calls for a detailed experimental inspection that may allow us to determine the location, width and type of the fundamental band gap of Sb$_2$Te$_3$. 

\begin{figure*}[t]
      \includegraphics[width=0.99\textwidth]{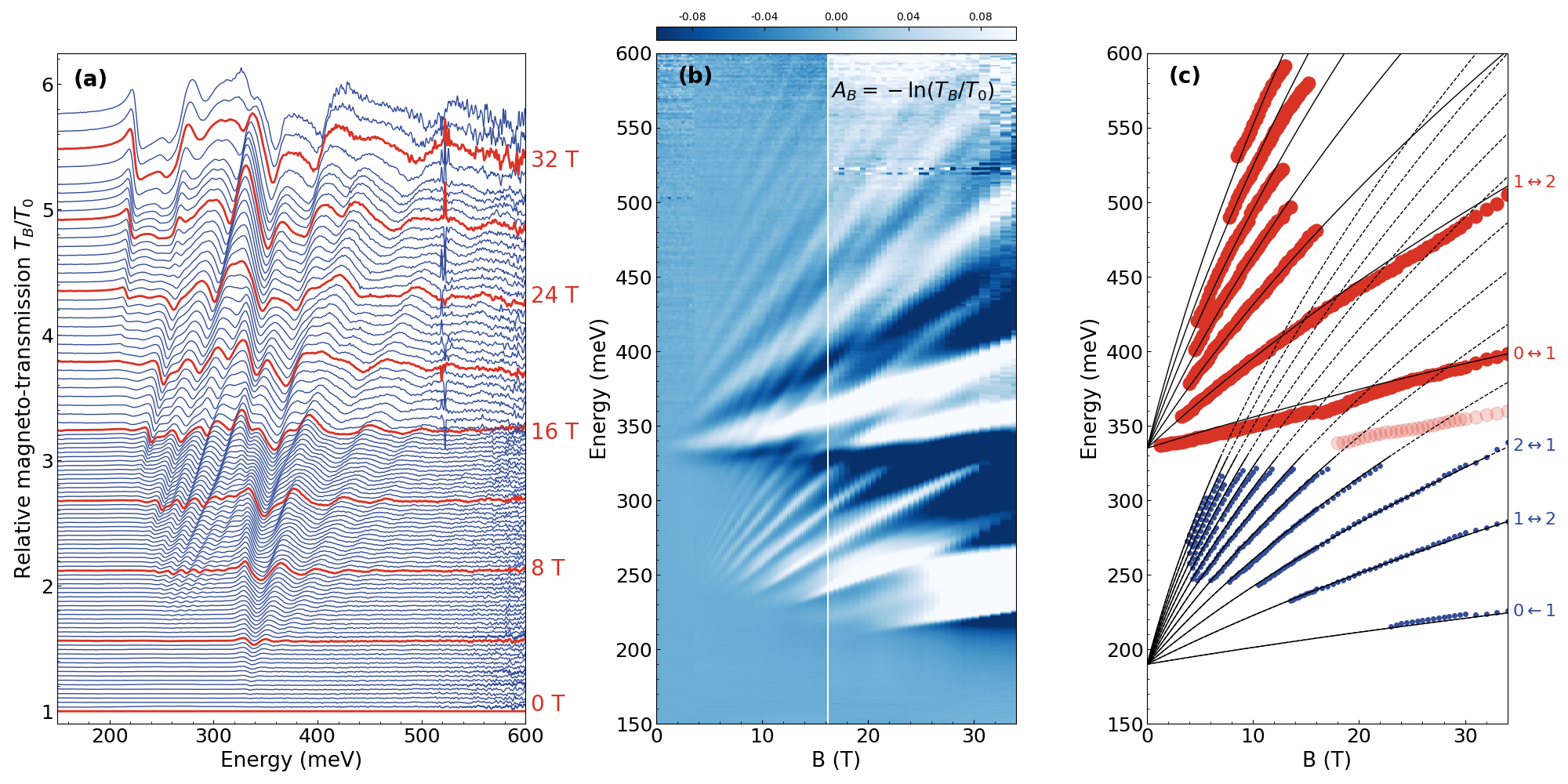}
      \caption{\label{Magneto-optics} Magneto-optical response of Sb$_2$Te$_3$ in the mid-infrared spectral range at $T=4.2$~K. (a) Relative magneto-transmission spectra, $T_B/T_0$, plotted at selected values of the applied magnetic field $B$. (b) A false-color plot of relative magneto-absorbance $A_B = -\ln[T_B/T_0]$. (c) Extracted positions of interband inter-LL resonances in the lower (blue) and upper (red) series. The light-red color is used for the satellite line discussed in the main text. The solid and dashed lines are fits using a simple two-band (massive Dirac-type) model, see the main text for a description.}
\end{figure*}

In this paper, we present a combined experimental and theoretical study of bulk electronic states in the Sb$_2$Te$_3$ topological insulator. Using magneto-optical, optical and magneto-transport methods, we find that Sb$_2$Te$_3$ is a material with a direct energy band gap. This gap reaches $E_g=(190\pm10)$~meV and it is located in the center of the BZ or along the trigonal axis. Nevertheless, the valence and conduction bands display also other local extrema at low energies, most likely in the mirror planes, and therefore, with a six-fold degeneracy. 

\section{Sample preparation, experimental and computational details}

The studied Sb$_2$Te$_3$ epilayer with a nominal thickness of $300$~nm (ellipsometry value: $276\pm2$~nm) was grown using molecular beam epitaxy on a 1-mm-thick (111)-oriented cleaved BaF$_2$ substrate. Compound Sb$_2$Te$_3$ and elemental Te sources were used for the control of the stoichiometry and composition. The deposition was carried out at a sample temperature of 290$^\circ$C at which a high-quality 2D growth mode occurs as evidenced by reflection high energy electron diffraction and atomic force microscopy. The grown Sb$_2$Te$_3$ layer has the trigonal axis oriented perpendicular to the substrate. 

The prepared epilayer was characterized using $x$-ray diffraction measurements in the temperature range of $T=$77-300~K. At room temperature, the deduced lattice constants, $a=(0.4257\pm0.0002)$~nm  and $c=(3.049\pm0.002)$~nm, compare well to values reported in literature~\cite{AndersonACSB74}: $a=0.4264$~nm and $c=3.0458$~nm, respectively. This suggests that the epilayer is well relaxed. The experiments also show that, when cooling down, a weak tensile stress (around 0.05\%) appears as a result of different thermal contractions in BaF$_2$ and Sb$_2$Te$_3$.   

%, or nearly relaxed. , with a possible weak lateral compressive stress around 0.1\%. Moreover, when cooling down, this compressive stress is partly released due to a higher thermal contraction of the BaF$_2$ substrate as compared to Sb$_2$Te$_3$.   

To measure infrared magneto-transmission, radiation from a globar or a mercury lamp was analyzed by a commercial Bruker Vertex 80v Fourier-transform spectrometer. The radiation was then delivered via light-pipe optics to the sample kept in the helium exchange gas at the temperature $T=2$ or 4.2~K and placed in a superconducting solenoid, or in the resistive high-field magnet (below or above 16 T, respectively), both at the LNCMI in Grenoble.

\begin{figure}
      \includegraphics[width=0.42\textwidth]{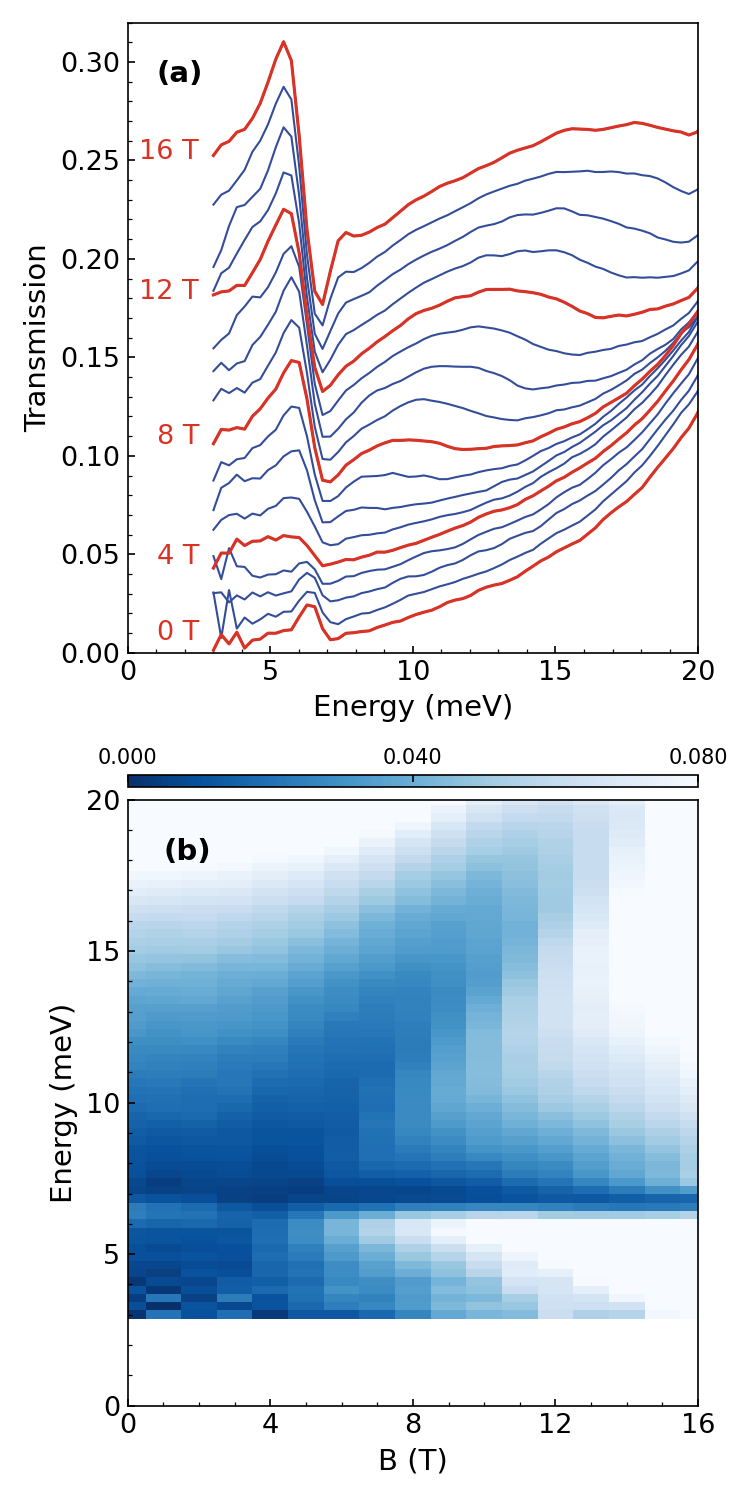}
      \caption{\label{CR} Magneto-optical response of Sb$_2$Te$_3$ in the far infrared spectral range at $T=4.2$~K. (a) Magneto-transmission spectra, $T_B$, normalized by the transmission of a bare BaF$_2$ substrate, are plotted for selected values of $B$. The same data in the form of a false-color plot are presented in (b). In both panels, pronounced coupling of the CR mode with the phonon mode at 7~meV is observed ($E_u$ mode~\cite{BragagliaSR20}).}
\end{figure}

Infrared radiation transmitted through the sample was detected by a composite bolometer, placed below the sample. The studied sample was probed in both Faraday and Voigt configurations, with the wave vector propagating along and perpendicular to the applied magnetic field, respectively. In the latter case, a linear polarizer was used to select the electric polarization oriented perpendicularly to the applied magnetic field.  The measured transmission spectra, $T_B$, were normalized by the zero-field transmission, $T_0$, and plotted in the form of relative magneto-transmission, $T_B/T_0$, or alternatively, as relative magneto-absorbance, $A_B=-\ln[T_B/T_0]$. 

The optical response at $B=0$ was deduced by the ellipsometry technique, with a commercial Woollam IR-VASE ellipsometer coupled to a closed He-cycle cryostat, for details see Ref.~\onlinecite{DubrokaPRB17}. Complementary magneto-transport measurements were realized using the standard lock-in technique on a sample with electrical contacts in the Hall-bar configuration, placed in the variable temperature insert in a superconducting coil (up to 16 T).  

The first-principles calculations were carried out with density functional theory (DFT) and the $GW$ approximation, as implemented in the codes FLEUR~\cite{fleur} and SPEX~\cite{FriedrichPRB10} within the FLAPW formalism (Full Potential Linearized Augmented Plane Waves). We used the same lattice parameters of Ref.~\cite{AguileraPRB19}. The DFT calculations were performed using the exchange-correlation functional GGA-PBE~\cite{PerdewPRL96}. We used an angular momentum cut-off $l_{max}$=10 and a plane-wave cutoff of 4.5~bohr$^{-1}$. The BZ was sampled using an $8\times 8 \times 8$ k-point mesh, both in the DFT and the $GW$ calculations. 

A mixed product basis was used for the $GW$ calculations, with an angular momentum cut-off of 5 and linear momentum cutoff of 2.9 bohr$^{-1}$. The number of bands was set to 500 for the calculation of the Green function and the polarization function. The spin-orbit coupling (SOC) effect was incorporated with the method of Refs. \cite{AguileraPRB13II,SakumaPRB11}, which ensured that the SOC effects were fully included in the Green function, the screened Coulomb potential, and the self-energy. 

The k-mesh used in the DFT and $GW$ calculations ensures~\cite{AguileraPRB19} that accurate Maximally Localized Wannier Functions (MLWFs)~\cite{MarzariRMP12} can be constructed. This was done with the help of the WANNIER90 library~\cite{MostofiCPC08}. Using the Wannier-interpolation technique as discussed in Ref.~\cite{AguileraPRB19}, one can obtain (interpolated) energy eigenvalues (DFT or $GW$) in a denser mesh. We have used it to interpolate the $GW$ eigenvalues in a $28\times28\times28$ mesh needed to converge the dielectric function. The latter was calculated from the RPA (Random Phase Approximation) polarizability~\cite{FriedrichPRB10} taking the macroscopic average including local-field effects.

\section{Results and Discussion}

\subsection{Magneto-optics}

\begin{figure*}
      \includegraphics[width=0.99\textwidth]{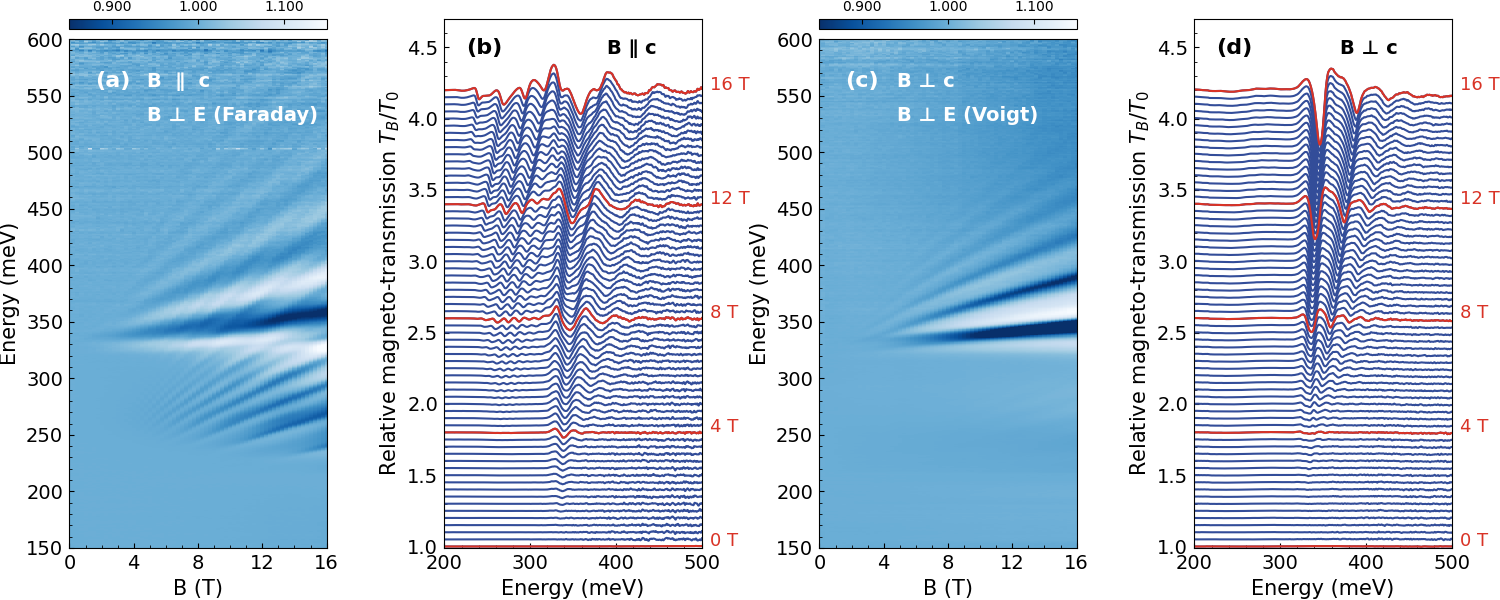}
      \caption{\label{Faraday-Voigt} Magneto-optical response of Sb$_2$Te$_3$ in the mid-infrared spectral range at $T=4.2$~K in the Faraday and Voigt (both $B\perp E$) configurations. Relative magneto-transmission spectra, $T_B/T_0$, collected in the Faraday and Voigt configurations are plotted in a form of false-color plots and stacked plots in (a,c) and (b,d), respectively.} 
\end{figure*}

The magneto-optical data collected on the explored Sb$_2$Te$_3$ epilayer at low temperatures in the mid-infrared spectral range are presented in Fig.~\ref{Magneto-optics}a and \ref{Magneto-optics}b as a stacked-plot of relative magneto-transmission spectra, $T_B/T_0$, and as a false-color plot of relative magneto-absorbance, $A_B$, respectively. The data are dominated by two series of resonances --  by the upper and lower set --  that may be directly associated with excitations bringing electrons across the band gap, between Landau-quantized valence bands. In the following, we analyze these magneto-optical data, and compare them with results of complementary techniques, in order to associate the two sets with particular locations in the BZ. In our magneto-optical data, we did not identify any contribution directly attributable to  surface states~\cite{HsiehPRL09,ZhuSR15,LundPCCP21,LocatelliSR22}.

Let us now analyze the observed transitions in terms of their position, $B$-dependence and intensity, in a greater detail. The transitions in both series follow a weakly sub-linear in $B$ dependence and extrapolate to a finite photon energy in the limit of a vanishing magnetic field. Such behaviour is typical of narrow-gap semiconductors and can be, in this particular case, well reproduced using a simple two-band Hamiltonian often referred to as the model of massive Dirac electrons. In this simple model, with two material parameters only, the conduction (+) and valence (-) bands display hyperbolic profiles: $E=\pm \sqrt{\Delta^2+\hbar^2v^2k^2}$, where $v$ is the velocity parameter and $2\Delta$ represents the width of the band gap. The magnetic field strong enough transforms these bands in dispersive one-dimensional Landau bands, with the band edge energies of $E_n=\pm \sqrt{\Delta^2+2 v^2 \hbar n |eB|}$, for $n\geq 0$. 

Taking the corresponding selection rules for electric-dipole excitations for massive Dirac electrons~\cite{LyJPCM16,LiuPRB10,OrlitaPRL15}: $n\rightarrow n \pm1$, the model reproduces the experimentally determined positions of interband inter-LL excitations very well, see Fig.~\ref{Magneto-optics}c. For the lower set, the parameters read: $2\Delta_L = (190\pm10)$~meV and $v_L = (4.1\pm0.2)\times10^5$~m/s. For the upper set, we obtained: $2\Delta_U = (340\pm10)$~meV and $v_U = (7.5\pm0.5)\times10^5$~m/s. The extracted parameters allow us to get estimates of the band-edge masses, $m_L=\Delta_L/v^2_L=(0.10\pm0.01)m_0$ and $m_U=\Delta_U/v^2_U=(0.055\pm0.005)m_0$, assuming that the electron-hole symmetry is preserved. $m_0$ stands for the bare electron mass. 

Let us note that the successful application of the simplified, massive Dirac, model is to a certain extent surprising. Clearly, we deal with a topological insulator, and the lower series of inter-LL excitation is -- as justified \emph{a posteriori} in the paper -- associated with the center of the BZ where the band structure is clearly inverted. In such a case, it is common to expand the Hamiltonian, to include a diagonal dispersive term $Mk^2$ which effectively describes the extension of the band inversion in the reciprocal space, see, \emph{e.g.}, Refs.~\cite{LiuPRB10,AssafSR16,LyJPCM16, TikuisisPRB21}.
This term impacts the LL spectrum mainly at high energies, \emph{i.e.}, well above the band gap. There, the LLs gain a linear in $B$ dependence, see Ref.~\cite{OrlitaPRL15} that contrasts with the $\sqrt{B}$ dependence typical of Dirac electrons (with $M\equiv 0$). In the case of Sb$_2$Te$_3$, we were only able to set the upper limit for the inversion parameter: $|M|<~\SI{10}{\electronvolt \cdot \AA^2}$. For higher values, the theoretical $B$-dependence of inter-LL excitations starts to deviate visibly from the experimental data. Another deviation from the simple massive Dirac model appears for the upper set in high magnetic fields. A satellite transition emerges at $B>15$~T below the $0 \leftrightarrow 1$ line, which disperses approximately with the same slope (light red full circles in Fig.~\ref{Magneto-optics}c). This additional line may correspond to an excitation from/to a shallow impurity state.

Other relevant pieces of information can be extracted from the intensity of the observed inter-LL excitations. Individual inter-LL transitions in the lower set emerge with increasing $B$ one by one, thus indicating a pronounced occupation effect. The quantum limit, with the Fermi energy in the lowest LL, is only achieved above 20~T when the transition $0 \leftrightarrow 1$ appears in the response ($\hbar\omega\approx 200$~meV, see Fig.~\ref{Magneto-optics}b). In fact, this is a well-known Moss-Burstein shift~\cite{BursteinPR54,MossPPS54}, but with the onset of interband absorption driven by the $B$-dependent Fermi energy. This implies that the lower set of transitions comes from a particular location in the BZ with a non-negligible density of free charge carriers. In contrast, no occupation effect is visibly manifested in the upper set. This indicates a lower, or even negligible, carrier density in the concerned bands. 

\begin{figure*}[t]
      \includegraphics[width=1.0\textwidth]{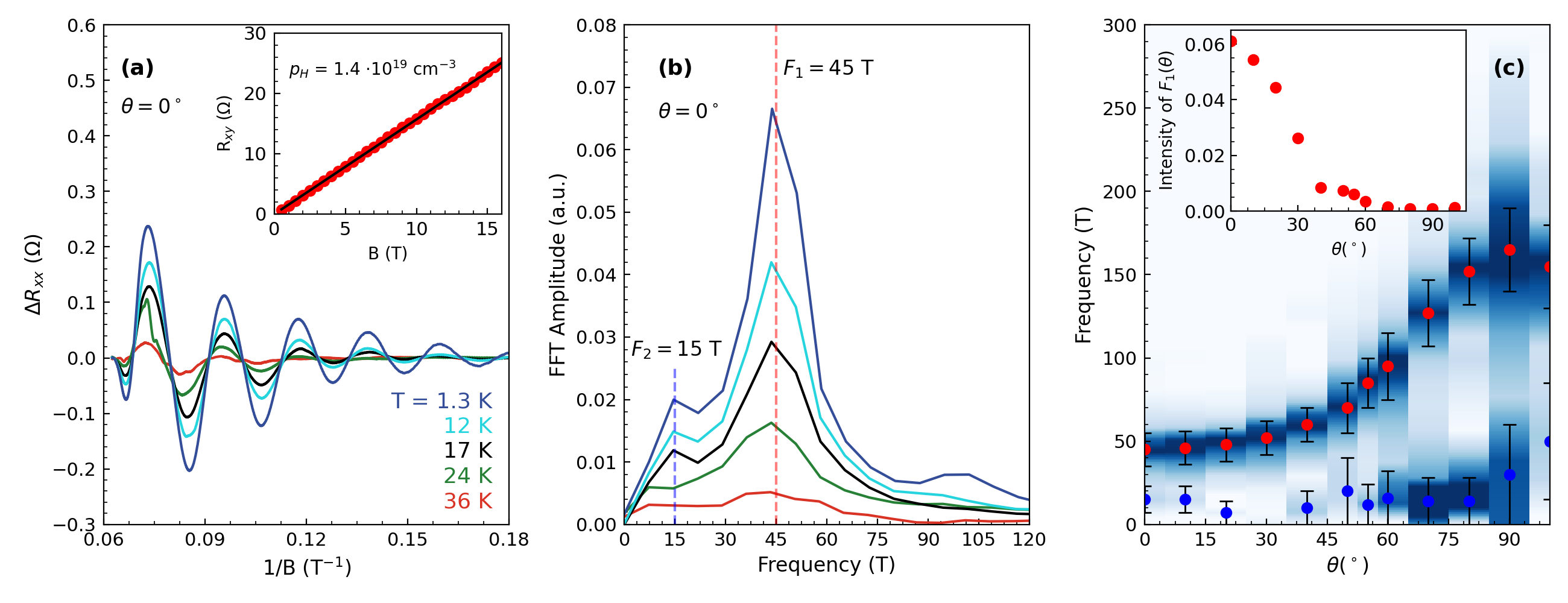}
      \caption{\label{Magneto-transport} Magneto-transport experiments. (a) Temperature dependence of background-removed Shubnikov-de Haas oscillations in Sb$_2$Te$_3$ with $B$ applied along the trigonal axis. The inset shows the corresponding Hall resistance at $T=1.36$~K, showing linear in $B$ dependence. (b) Frequency spectra of $1/B$-periodic Shubnikov-de Haas oscillations at selected temperatures. (c) False-color plot: frequency spectra of $1/B$-periodic Shubnikov-de Haas oscillations as a function of the angle $\theta$ between the trigonal axis and magnetic field. To visualize the angle dependence, individual experimental traces were normalised by their maximum. The angle-dependence of the oscillations' amplitude is plotted in the inset.}
\end{figure*}

Magneto-optical experiments at lower photon energies, below the reststrahlenband of the BaF$_2$ substrate, were performed as well, see Fig.~\ref{CR}. As expected, the transmission of the sample is low, due to absorption on free charge carriers (Drude-type absorption). This takes a form of cyclotron resonance (CR) absorption when the magnetic field is applied. Unfortunately, the precise read-out of the CR energy and the extraction of the effective mass is not straightforward due to strong coupling, primarily due to a dielectric effect, between a pronounced phonon mode ($E_u\approx 7$~meV, see Ref.~\cite{BragagliaSR20}) and free charge carriers. Nevertheless, a rough estimate of the effective mass $m_c\approx 0.08m_0$ from the $B$-dependent minimum in the transmission response approaches the Dirac mass $m_L$ deduced for the lower set of interband excitations. 

To probe the anisotropy of the magneto-optical response, a complementary magneto-optical experiment was carried out with $B$ oriented perpendicular to the trigonal axis of Sb$_2$Te$_3$ ($B\perp c$). In the case of a thin layer, such an experiment is only possible in the Voigt configuration. To follow transitions active within the same set of selection rules as in the Faraday configuration, we used a linear polarizer and selected radiation with the electric field component perpendicular to the magnetic field ($B\perp E$). The data collected in the Faraday and Voigt configuration are compared in Fig.~\ref{Faraday-Voigt}.

Interestingly, the lower set of inter-LL excitations vanishes completely in the Voigt configuration ($B\perp c$). This suggests a pronounced anisotropy of the band structure at the related location in the BZ, with the bands too flat along the $c$ direction to enter the Landau-quantized regime at magnetic fields applied. In contrast, the upper series of transitions shows nearly isotropic behaviour. Again, the observed response (\emph{i.e.}, the upper set) can be well reproduced using the model of massive Dirac electrons, with an identical band gap $2\Delta_U$, but with the velocity parameter slightly lowered, $v_U=(6.0\pm0.5)$~m/s. Let us note that two independent experiments were realized in the Voigt configuration, with the sample rotated by 90 deg around the trigonal axis to probe the in-plane anisotropy. These measurements provide us with nearly identical results, consistent with a fairly low in-plane anisotropy (weak trigonal warping).   

To sum up, the magneto-optical response of Sb$_2$Te$_3$ at low photon energies is surprisingly complex. It is dominated by two, or when the expected multiplicity of the band extrema is considered, more than two distinct locations in the BZ. These differ considerably in the extracted band gap, by the characteristic anisotropy of the local band structure, but also by the density of free charge carriers at the local band extrema. The observed magneto-optical response suggests that Sb$_2$Te$_3$ is a direct-gap semiconductor with the band gap reaching $E_g=(190\pm10)$~meV. 

\subsection{Magneto-transport}

To complement our magneto-optical experiments, we performed a series of magneto-transport measurements on the same sample. Both transversal and longitudinal components of the magneto-resistivity tensor -- with $B$ applied along the trigonal axis of Sb$_2$Te$_3$ --  were recorded, the latter also as a function of temperature, see Fig.~\ref{Magneto-transport}a. The Hall signal was linear in the applied magnetic field and the sign indicated $p$-type conductivity typical of Sb$_2$Te$_3$~\cite{SchwartzSSC67,Kohlerpss77,KulbachinskiiPRB95,ZhaoPRB19,KulbachinskiiMTP21}. The slope corresponds to the total hole density of $p_H=(1.4\pm 0.1)\times 10^{19}$~cm$^{-3}$. The presence of this non-negligible hole concentration in single crystalline bulk or epitaxial layers of Sb$_2$Te$_3$ may be caused by the rather similar cation and anion electronegativities~\cite{CavaJMCC13} in this compound. During crystal growth, this leads to a large number of negatively charged antisite defects which result from the occupation of some Te sites in the Sb$_2$Te$_3$ lattice by Sb and which produce one hole for one antisite defect~\cite{CavaJMCC13,Horakpssa95,RajputCGD23}, and furthermore also to the formation of Sb vacancies~\cite{RajputCGD23}.

The longitudinal component of magneto-resistance exhibited well-visible Shubnikov-de Haas oscillations. Our analysis based the fast Fourier transformation indicated two oscillation periods, having frequencies of $F_1=(45\pm5)$~T and $F_2=(15\pm5)$~T, see Fig.~\ref{Magneto-transport}b. This indicates the existence of two types of hole Fermi surfaces in our sample. Below, we refer to them as larger and smaller ones, respectively. Interestingly, no signs of splitting due to spin was observed. This may be surprising in a material with a particularly strong spin-orbit interaction. Nevertheless, in the sister material Bi$_2$Se$_3$, the same effect is present as well and explained as due to spin splitting that matches multiples of the cyclotron energy~\cite{Kohlerpssb75}. The Lifshitz-Kosevitch-like analysis of oscillations' damping with temperature was only possible for the dominant oscillations, corresponding to the larger Fermi surface. It provided us with the effective mass of $m_1 = (0.11\pm0.02)m_0$. When the approximation of a strictly parabolic band is used, this gives us an estimate of the Fermi energy around 40-50~meV. 

To explore the anisotropy of the Fermi surfaces, we have also traced the Shubnikov-de Haas oscillations as a function of the angle $\theta$ between the trigonal axis and the direction of the applied magnetic field ($\theta=0$ equiv. $B \perp c$). In our experiment, the magnetic field was oriented randomly with respect to the in-plane ($a$-$b$) crystallographic axes. The result of the frequency analysis is plotted in a form of a false-color plot in Fig.~\ref{Magneto-transport}c. 

Interestingly, the angle dependence is distinctively different for the two identified oscillation periods. The frequency $F_1$, belonging to the larger Fermi surface, increases monotonically with the angle and reaches roughly 4$\times$ its original value at $\theta=90$~deg. This excludes the possibility that we observe the response of electrons in the surface states, recently reported on even thinner Sb$_2$Te$_3$ layers~\cite{WeyrichAPL17}. The amplitude of oscillations drops significantly, see the inset of Fig.~\ref{Magneto-transport}c, becoming very weak at angles around $\theta=90$~deg. The larger Fermi surface is thus greatly elongated in the direction of the trigonal axis. The frequency $F_2$ -- corresponding to the smaller Fermi surface -- remains constant as a function of $\theta$ within the experimental accuracy, and hence, this Fermi surface is approximately spherical. 

Let us note that a different angle dependence was observed in early magneto-transport studies of Sb$_2$Te$_3$. The observed oscillation period did not evolve monotonically with $\theta$~\cite{SchwartzSSC67} and even signs of splitting appeared when $B$ was not parallel with the trigonal axis~\cite{MiddendorffSSC73}. Such behavior was interpreted as due to the multiple-valley structure of the topmost valence band (with the valley degeneracy $N=6$ or 12). It is important to stress, however, that those studies were performed on crystals with almost an order of magnitude larger hole density (around 10$^{20}$~cm$^{-3}$). 

Having quantified the anisotropy of the larger Fermi surface, we may estimate the related per-valley density of holes. Let us assume that the band, even though profoundly anisotropic, has still a strictly parabolic profile at the concerned energies. The observed angle dependence of $F_1$ implies a relatively large difference -- by a factor of 16 -- between the effective masses perpendicular to and along the trigonal axis: $m^\perp_1=(0.11\pm0.02)m_0$ and $m^\|_1=(1.8\pm0.3)m_0$, respectively. This allows us to estimate, taking the frequency $F_1 = 45$~T at $\theta=0$, the corresponding hole density per valley: $p_1 = (m_1^\|/m_1^\perp)^{1/2} (2eF_1/\hbar)^{3/2}/(3\pi^2)=7\times10^{18}$~cm$^{-3}$. 

The extracted per-valley hole density $p_1$ represents a half of the total hole density $p_H$ estimated in the Hall experiments. Even though there exists some uncertainty in the Hall read-out of the carrier density in systems with two types of charge carriers, this suggests that the larger hole pocket might be located at the $\Gamma$ point ($N=1$). Alternatively, the double valley degeneracy, $N=2$, with two larger hole pockets located along the trigonal axis (along the $Z-\Gamma-Z$ line) is also possible. A higher valley degeneracy, $N=6$ or even $N=12$, is not consistent with our findings since $N \cdot p_1\gg p_H$. Independently of the number of valleys concluded, each one is always considered to be twice degenerate due to spin.   

The extracted properties of the larger hole pocket, \emph{i.e.}, the anisotropy, effective mass, and the Fermi energy, remarkably resemble the analogous parameters deduced for the lower set of interband inter-LL transitions in Fig.~\ref{Magneto-optics}. Hence, it is plausible to assume that they have their origin in the very same location of the BZ. This implies that the fundamental direct band gap of Sb$_2$Te$_3$ is located at the $\Gamma$ point ($N=1$) or alternatively along the $Z-\Gamma-Z$ line ($N=2$), cf. Fig.~\ref{BZ}.  

Assigning the larger Fermi surface, observed in magneto-transport experiments, to the center of the BZ, one may speculate about the location(s) of the smaller one(s). It is plausible to assume that they may be positioned away from the trigonal axis  -- on the binary axis, in the mirror place (both $N=6$) or in a general location lacking any higher symmetry ($N=12$). Most likely, those were the hole pockets which dominated the response in early quantum oscillation experiments performed on bulk samples with a significantly higher hole density~\cite{SchwartzSSC67,MiddendorffSSC73}. 

\begin{figure}[t]
      \includegraphics[width=0.48\textwidth]{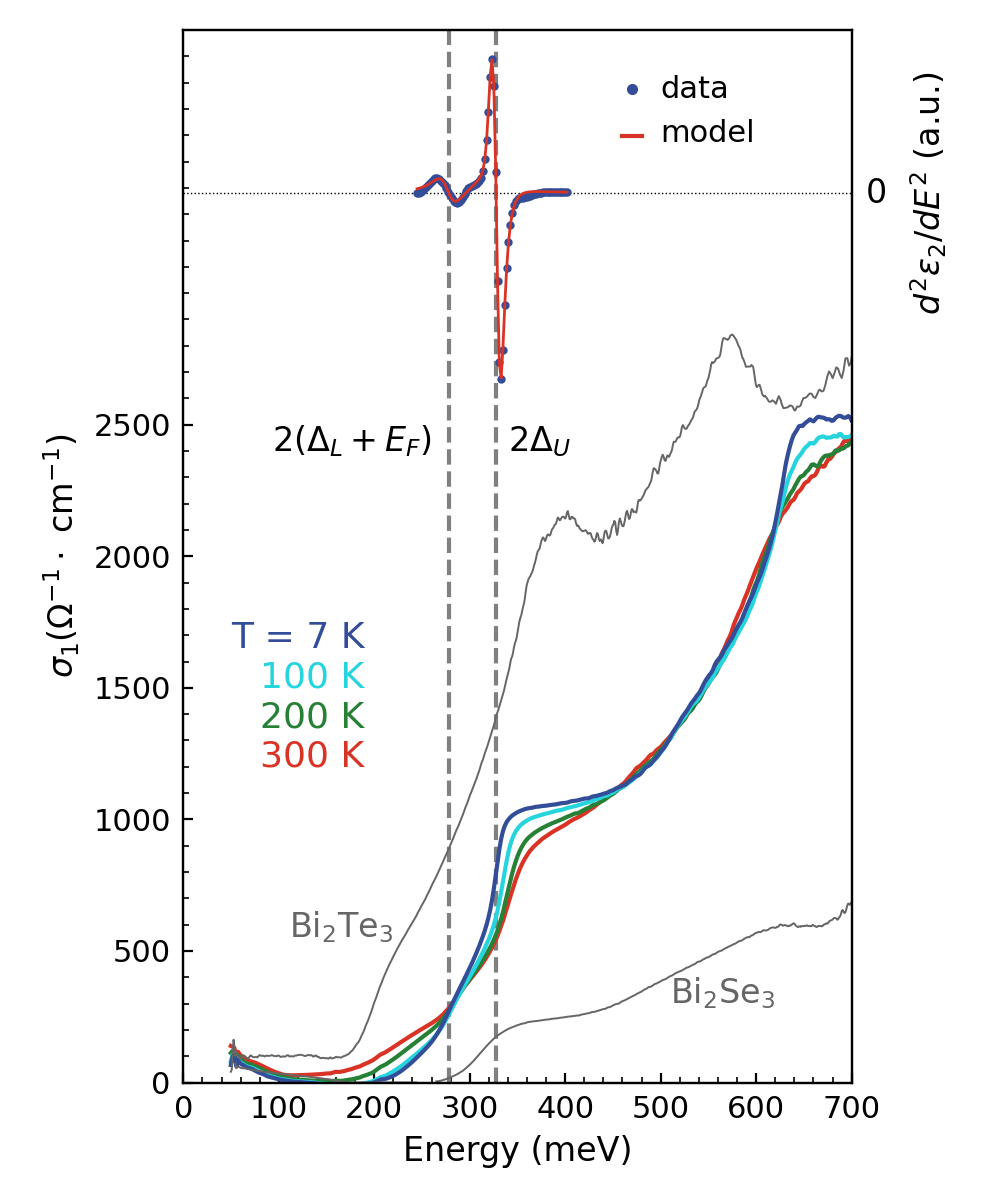}
      \caption{\label{Zero-field} Real part of optical conductivity of Sb$_2$Te$_3$ deduced using ellipsometry at $T=7, 100, 200$ and 300~K. The vertical dashed lines mark the position of two inflection points (at $T=7$~K) in the imaginary part of the dielectric function, found as zero-points of the second derivative, see the upper part. They are associated with two onsets of interband absorption in the data. The gray curves for Bi$_2$Se$_3$ and Bi$_2$Te$_3$ are re-plotted from Refs.~\cite{DubrokaPRB17} and \cite{MohelskyPRB20}, respectively.}
\end{figure}

\subsection{Ellipsometry measurements}

Let us now confront our conclusions so far with the result of the ellipsometry experiments. These were performed at $B=0$ and at varying $T$: from room temperature down to 7~K. The deduced real part of optical conductivity, the optical constant most relevant for this work, is plotted in Fig.~\ref{Zero-field}. It has a double-step onset at low photon energies, which smears out with the increasing temperature. To estimate the positions of these two onsets, we used the critical point model~\cite{YuFS96} and searched for zero-points in the second derivative of the imaginary part of the dielectric function (plotted in the top part of Fig.~\ref{Zero-field}). The positions of onsets are marked by two vertical dashed lines at 277 and 329~meV in Fig.~\ref{Zero-field}. 

The lower inflection point matches well the expected onset of interband absorption at the $\Gamma$ point (or along the trigonal axis), corresponding to the fundamental band gap enhanced by twice the Fermi energy at the larger hole pocket: $2(\Delta_L+E_F)$. This is due to the above mentioned Moss-Burstein shift~\cite{BursteinPR54,MossPPS54} typical of degenerate semiconductors with a low electron-hole asymmetry. The upper inflection point corresponds well to the onset of interband absorption at the energy of the local band gap $2\Delta_U$. It is worth noting that,
apart from smearing clearly visible in the spectra, the basic character of the zero-field optical response does not change with an increasing temperature. This implies that the band structure is robust against small temperature-induced strain/stress observed in the $x$-ray diffraction experiments.  

It is instructive to compare the zero-field optical response of Sb$_2$Te$_3$ with sibling materials: Bi$_2$Se$_3$ and Bi$_2$Te$_3$, in which the valley degeneracy of the fundamental band gap is consensually established: $N=1$ and $N=6$, respectively. Interestingly, the profile of interband absorption in Bi$_2$Se$_3$ resembles -- by its shape and amplitude -- the first step in the absorption of Sb$_2$Te$_3$, just shifted towards higher energies. On the other hand, at higher photon energies, the optical conductivity of Bi$_2$Te$_3$ has the profile and overall strength similar to Sb$_2$Te$_3$, only shifted towards lower energies. 

The above considerations are qualitative only, but they are in line with the above concluded single, or at most, double valley degeneracy of the fundamental direct band gap of $2\Delta_L=190$~meV in Sb$_2$Te$_3$. In addition, they point towards the sixfold valley degeneracy of the upper (local) band gap of $2\Delta_U=340$~meV. We suggest that this local band gap can be associated with the smaller hole pocket (the frequency $F_2$) observed in magneto-transport experiments.

\subsection{Comparison with output of $GW$ calculations}

\begin{figure*}[t]
      \includegraphics[width=0.98\textwidth]{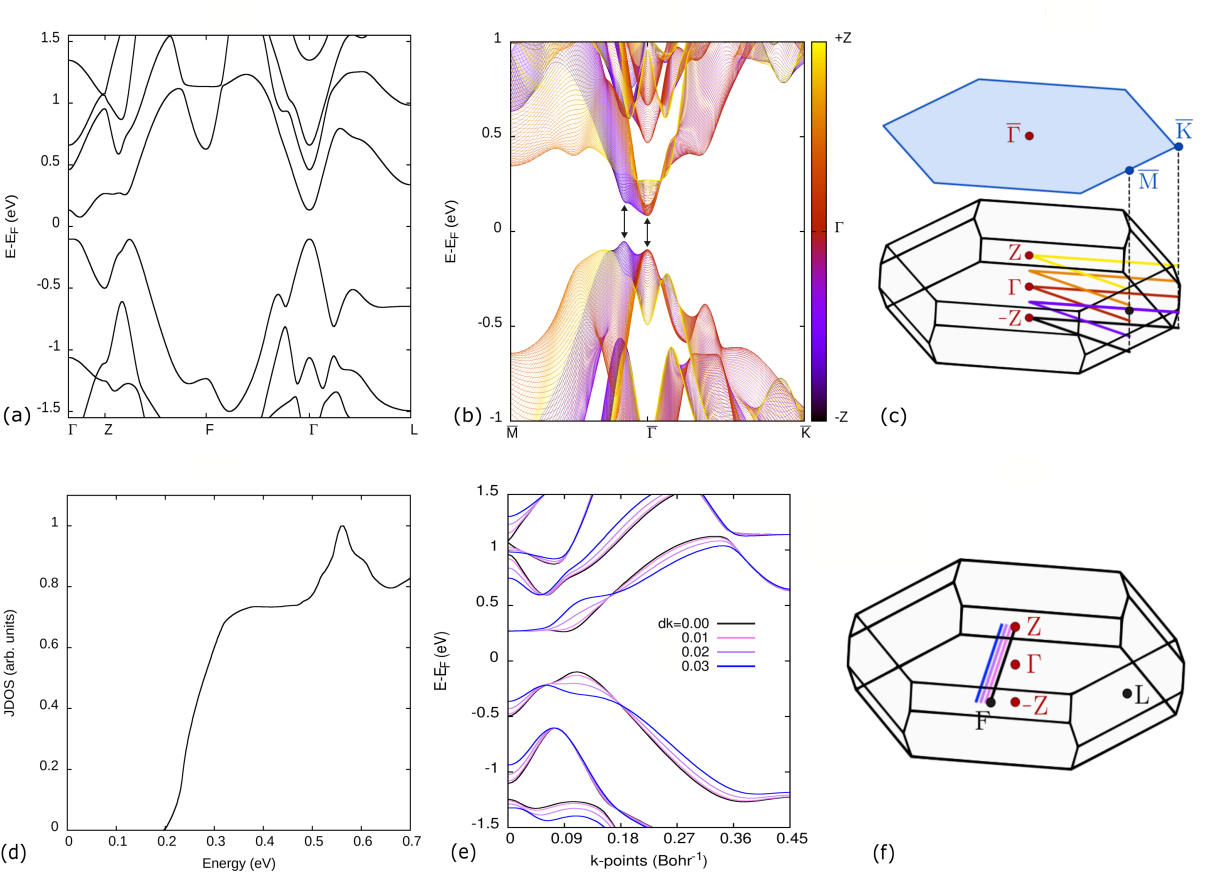}
      \caption{\label{combined-theory} Band structure of Sb$_2$Te$_3$ calculated using the $GW$ method along different paths of the BZ: (a) Standard high-symmetry path of the bulk BZ. (b) Bulk bands projected on the surface (2D) BZ. Each line, corresponds to a path in the bulk BZ that projects onto the path $\bar{{\rm{M}}}-\bar{\Gamma}-\bar{{\rm{K}}}$ in the surface BZ. The color of the bands represents the $k_z$ component of the path. The paths and the corresponding colors are indicated in panel (c). 
      Panel (d) shows the calculated joint density of states at low photon energies.
      Panel (e): bands along the Z$-$F path of the bulk (blue) and three paths slightly shifted with a fixed $k_z$. This shows that the maximum and minimum are indeed located along the Z$-$F line and have therefore six-fold degeneracy ($N=6$. The paths and colors corresponding to (e) are shown in panel (f). }
\end{figure*}

Let us now compare our inferences -- made so far solely via the analysis of the collected experimental data -- with the electronic band structure calculated using the $GW$ approach. The band structure obtained theoretically along the selected directions of the BZ is plotted in Fig.~\ref{combined-theory}a. A series of band structure cuts was also depicted in Fig.~\ref{combined-theory}b. These were made for selected values of $k_z$ along particular directions in the reciprocal space, see Fig.~\ref{combined-theory}c,  
and finally, projected to the surface (2D) BZ.   

In the big picture, the calculated band structure is in a very good agreement with our conclusions based solely on the experimental data. The extrema of electronic bands relevant for low-energy excitations targeted in our experiments are indeed located at the $\Gamma$ point, along the $Z-\Gamma-Z$ line and in the mirror planes. The calculations suggest the existence of two nearly degenerate maxima of the topmost valence band. This is consistent with the appearance of two series of interband inter-LL transitions in the magneto-optical response (Fig.~\ref{Magneto-optics}) as well as two onsets in the zero-field optical conductivity (Fig.~\ref{Zero-field}). The vertical double arrows in Fig.~\ref{combined-theory}b indicate corresponding locations in the BZ. The predicted band gap is slightly below 200~meV, a value perfectly in line with our optical and magneto-optical data ($2\Delta_L$). The calculations also suggest fairly large anisotropy (in the direction along versus perpendicular to the trigonal axis) of the global valence band maximum at the $\Gamma$ point.  In addition, the calculated step-like nature of the joint density of states (Fig.~\ref{combined-theory}d) compares rather well with the onset of interband absorption visualized by ellipsometry experiments (Fig.~\ref{Zero-field}). 

Having said this, let us compare the experimental findings with theoretical predictions in a greater detail, to identify the existing differences. While our transport experiments on the $p$-type sample conclude at most a double degeneracy of the global valence-band maximum, a sixfold degeneracy is suggested by calculations (Fig.~\ref{combined-theory}b). This discrepancy is also related to another difference: The calculations imply an indirect type of the band gap, while the experimental data are rather consistent with a direct one. Both discrepancies have the same origin: the theoretical prediction of the two maxima in the valence band with similar energies. The highest one (the absolute valence band maximum) lies on the mirror plane (degeneracy $N=6$). To confirm this, Figs.~\ref{combined-theory}e and f show the band structure for three other off-symmetry lines in the BZ parallel to the Z$-$F line (\emph{i.e.}, parallel to the mirror plane). All paths have the same $k_z$ component and deviate from each other by 0.01 Bohr$^{-1}$ in-plane. The bands show that the VBM lies indeed along the Z$-$F line, and is thus six-fold degenerate ($N=6$). The second one, approximately 40~meV below, is at the $\Gamma$ point (degeneracy $N=1$). However, as we discussed in Ref.~\cite{NechaevPRB15}, the relative energy between these two maxima is extremely sensitive to the details of the calculations. For example, a different set of lattice parameters used in the calculation can shift the valence band maximum from one position to the other (see Fig.~2 of Ref.~\cite{NechaevPRB15}). Therefore, in this case, the band structure calculations cannot give a definitive answer about which of the two is the absolute maximum.  

It is interesting to compare these findings to the band structure of the sister compound Bi$_2$Te$_3$. There, a consensus exists in literature, both in experimental studies~\cite{Kohlerpssb76II,MohelskyPRB20} and in theoretical calculations, that the valence band is characterized by six pronounced global maxima located in the mirror planes ($N=6$). The results of $GW$ calculations performed for the mixed compound (Sb$_{1-x}$Bi$_{x})_2$Te$_3$ show such a trend when the Sb-to-Bi ratio increases. This is shown in Fig.~\ref{mixed} in the Appendix, where we also discuss the role of the spin-orbit interaction in this cross over (Fig.~\ref{SOC}). Namely, the gradual increase in the bismuth concentration suppresses the valence band maximum at the $\Gamma$ point and enhances the maxima in the mirror planes ($Z$-$F$ direction). These latter maxima become the global ones for $x>0.6$. This picture is also consistent with conclusions of early quantum-oscillation experiments performed on the mixed (Sb$_{1-x}$Bi$_{x})_2$Te$_3$ crystals~\cite{Kohlerpss77}. The comparison of band structure parameters deduced in this work for Sb$_2$Te$_3$ with those for Bi$_2$Te$_3$~\cite{MohelskyPRB20} is made in Tab.~\ref{Tab}.

\begin{table*}
    \centering
    \renewcommand{\arraystretch}{1.25}
    \begin{tabular}{|x{25mm}|x{25mm}|x{35mm}|x{25mm}|x{40mm}|}
 \hline
     \multirow{2}{*}{\textbf{Material}} & \textbf{Direct band} & \textbf{Location in} & \textbf{Valley} & \textbf{Effective band-edge}\\
       & \textbf{gap (meV)} & \textbf{Brillouin zone} & \textbf{degeneracy} & \textbf{mass} $(m_0)$\\
\hline
\hline
     \multirow{3}{*}{Sb$_2$Te$_3$} & \multirow{2}{*}{$190\pm10$} & \multirow{2}{*}{$\Gamma$ (or $Z-\Gamma-Z$)} & \multirow{2}{*}{1 (or 2)}& $a-b$ plane: $0.11\pm0.02$\\
                                   &                             &   &  & $c$ axis: $1.8\pm0.3$ \\ 
                                   & $340\pm10$ & Mirror planes & 6 & $0.055\pm0.005$ \\
     
\hline
Bi$_2$Te$_3$~\cite{MohelskyPRB20} & $175\pm10$ & Mirror planes & 6 & $0.070\pm0.005$\\
              
\hline  
    \end{tabular}
    \caption{Summary of band structure parameters deduced for Sb$_2$Te$_3$ in this work compared to those of Bi$_2$Te$_3$ taken from Ref.~\cite{MohelskyPRB20}. The reported band edge mass corresponds to both electrons and holes, thus neglecting completely a possible electron-hole asymmetry.}
    \label{Tab}
\end{table*}

Finally, we have experimentally shown that, besides the fundamental band gap in the centre of the BZ, the band structure comprises also additional, local energy band gaps, located most likely in the mirror planes and with fairly high, in their vicinity, isotropic bands. However, these additional band gaps are not clearly identified in our calculations. Even though, such a maximum of the calculated topmost valence band can be found in the mirror plane, the shape of rather shallow counterpart in the conduction band would not allow to form a well-defined Landau quantization at the scale seen in our experimental data. 

\section{Conclusions}

We conclude that antimony telluride is a semiconductor with a direct energy band gap of $E_g = (190 \pm 10)$~meV at low temperatures, which is located at the center of the BZ, or alternatively, along the trigonal axis. In the observed response, we do also identify additional local maxima of the valence band, nearly degenerate with the global one, which form additional (local) direct gaps with the width of
$(340\pm10)$~meV and displaying the six-fold valley degeneracy (mirror planes). Our findings are in fairly good agreement with the electronic band structure calculated using the $GW$ method, thus demonstrating high predictive power of this advanced theoretical technique. 

\begin{acknowledgments}
The authors also acknowledge the support of the LNCMI-CNRS in Grenoble, a member of the European Magnetic Field Laboratory (EMFL).  The work has been supported by the ANR projects DIRAC3D (ANR-17-CE30-0023) and COLECTOR (ANR-19-CE30-0032). This research was supported by the NCCR MARVEL, a National Centre of Competence in Research, funded by the Swiss National Science Foundation (grant number 205602). IA and MFT acknowledge the computing time granted through SURF on the Dutch National Supercomputer Snellius. X.S. and B.P. acknowledge support from ANR Grant No. ANR-20-CE30-0015-01. We acknowledge the support by the project Quantum materials for applications in sustainable technologies, CZ.02.01.01/00/22\_008/0004572, the Czech Science Foundation (GACR) under Project No. GA20-10377S and the CzechNanoLab Research Infrastructure supported by MEYS CR (LM2023051). For the first-principles calculations, we used the Dutch national e-infrastructure with the support of the SURF Cooperative using grant no. EINF-5380. 
\end{acknowledgments}

\appendix*

\section{Electronic band structure of (Sb$_{1-x}$Bi$_{x})_2$Te$_3$}

\begin{figure*}
      \includegraphics[width=0.77\textwidth]{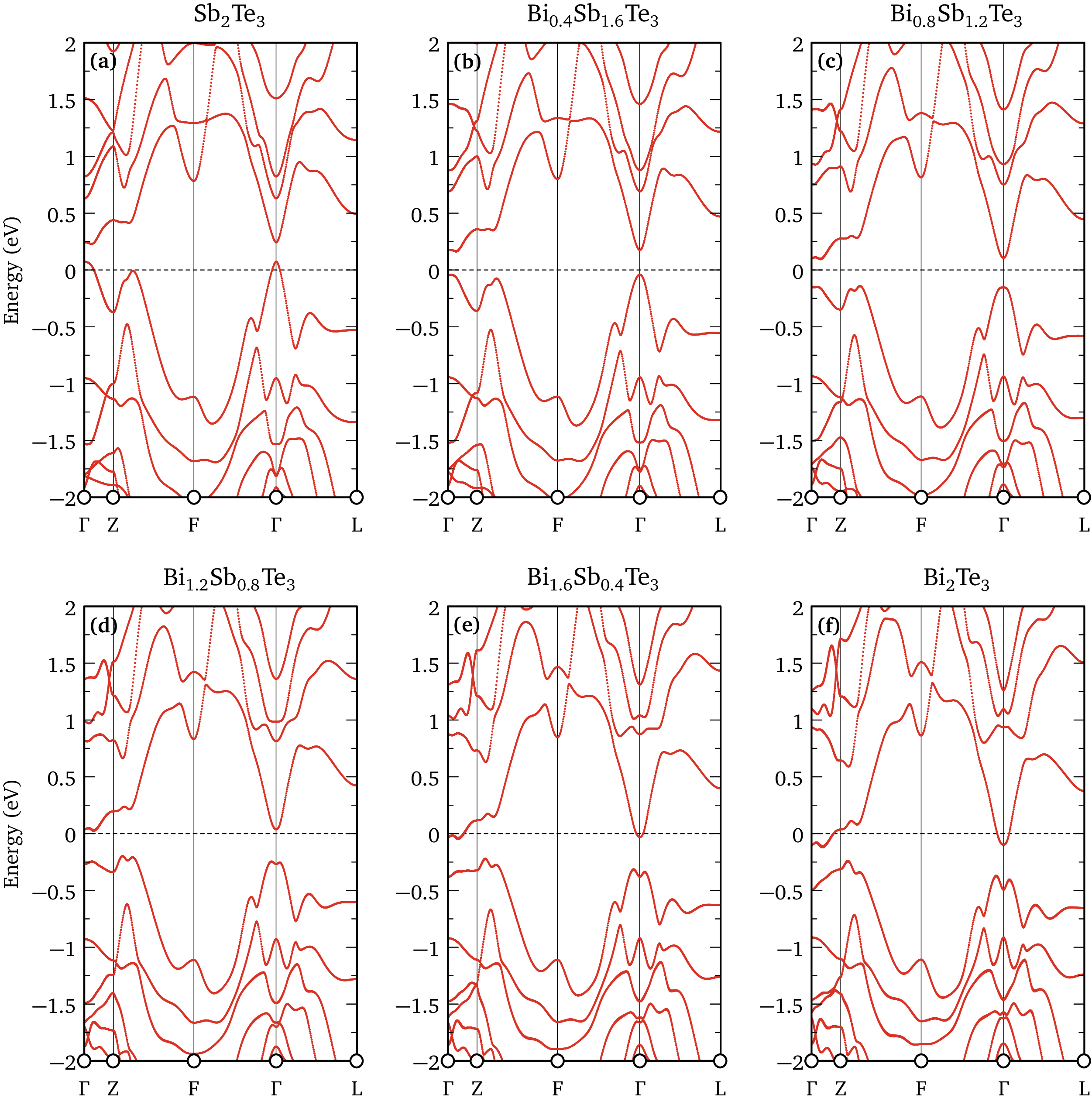}
      \caption{\label{mixed} The electronic band structure of (Sb$_{1-x}$Bi$_{x})_2$Te$_3$ calculated using $GW$ method along selected directions for the bismuth content: $x=0, 0.2, 0.4, 0.6, 0.8$ and 1.}
\end{figure*}

Following the approach described in Ref.~\cite{AguileraPRB19}, we can use the GW calculations to construct tight-binding Hamiltonians for the parent compounds Sb$_2$Te$_3$ and Bi$_2$Te$_3$, that will allow us to simulate alloys with intermediate Sb and Bi concentrations. These tight-binding Hamiltonians are expressed in a basis of Wannier functions and there are no adjustable parameters: the Hamiltonian matrix elements are fully determined by the GW calculation of these two compounds. We used a 6$\times$6$\times$6 k-point mesh for these calculations. A doping was simulated for the parent compounds by fitting the calculated band extrema to the ARPES results from~\cite{EschbachNC15}. To simulate the band structures of the alloys in Fig.~\ref{mixed}, the tight-binding parameters are smoothly varied between those of the parent compounds using the virtual crystal approximation (VCA). 

\begin{figure}
      \includegraphics[width=0.37\textwidth,angle=270]{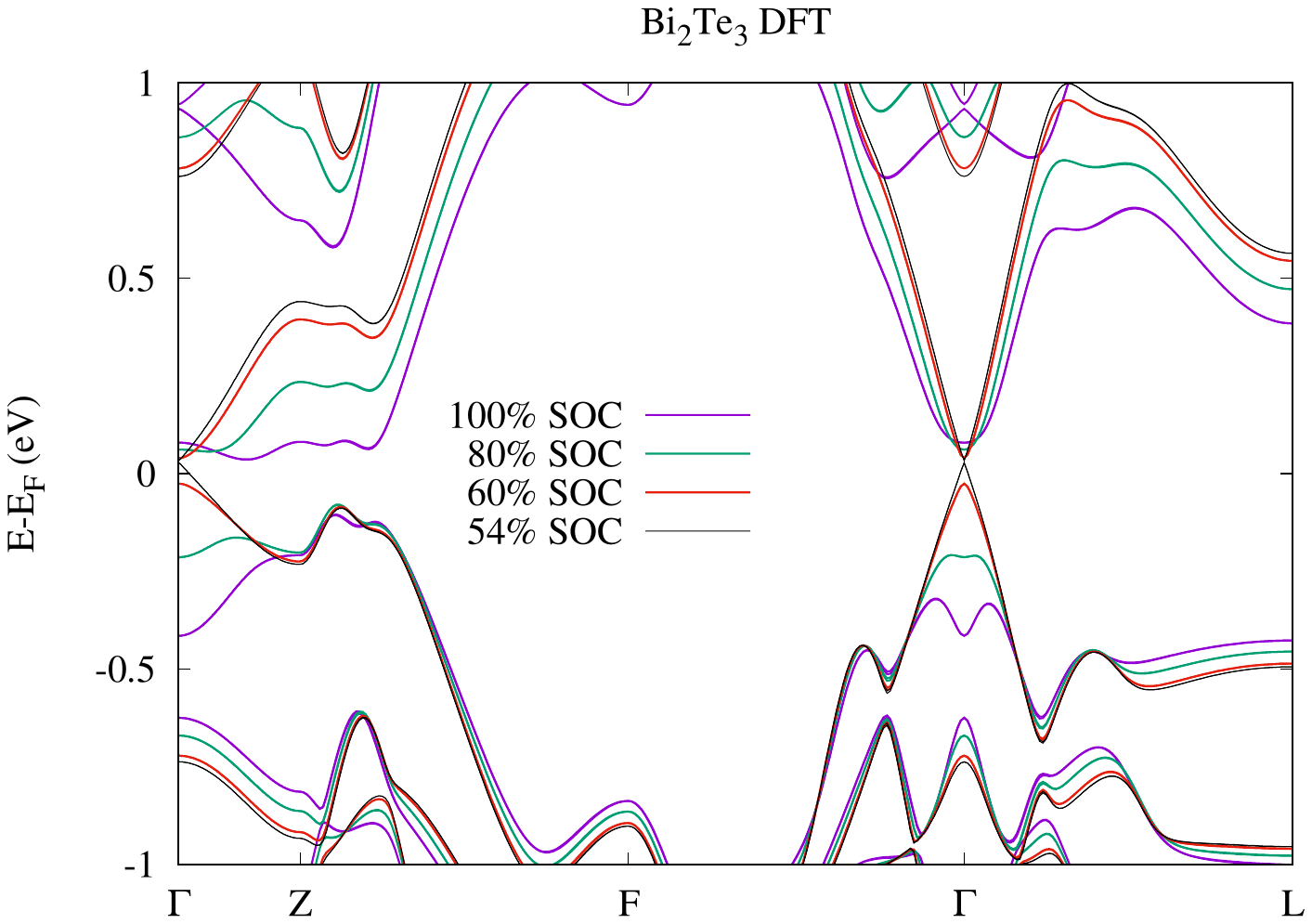}
      \caption{\label{SOC} DFT calculations in the generalized gradient approximation (GGA) of Bi$_2$Te$_3$ performed with the FLAPW code FLEUR~\cite{fleur}. The SOC strength is reduced from 100\% to 54\%. After that, the system undergoes a topological phase transition and becomes trivial.}
\end{figure}

It is worth to note that the qualitative change of the Sb$_{1-x}$Bi$_{x})_2$Te$_3$ band structure, from one to six maxima in the valence band, when Sb is gradually replaced by Bi, can be explained in terms of the increasing strength of the spin-orbit interaction that is significantly stronger for the latter element. To elucidate this, we present in Fig.~\ref{SOC} DFT calculations of Bi$_2$Te$_3$ in which we artificially reduced the spin-orbit strength from 100\% down to 54\% corresponding to the point of a topological phase transition. For lower spin-orbit interaction strengths, the system becomes topologically trivial. The results of this analysis are shown in Fig.~\ref{SOC} and suggests that the existence of the two competing valence band maxima in Sb$_2$Te$_3$ are indeed a consequence of the lower SOC strength in Sb$_2$Te$_3$ with respect to Bi$_2$Te$_3$.

\bibliography{Sb2Te3}

\begin{thebibliography}{49}%
\makeatletter
\providecommand \@ifxundefined [1]{%
 \@ifx{#1\undefined}
}%
\providecommand \@ifnum [1]{%
 \ifnum #1\expandafter \@firstoftwo
 \else \expandafter \@secondoftwo
 \fi
}%
\providecommand \@ifx [1]{%
 \ifx #1\expandafter \@firstoftwo
 \else \expandafter \@secondoftwo
 \fi
}%
\providecommand \natexlab [1]{#1}%
\providecommand \enquote  [1]{``#1''}%
\providecommand \bibnamefont  [1]{#1}%
\providecommand \bibfnamefont [1]{#1}%
\providecommand \citenamefont [1]{#1}%
\providecommand \href@noop [0]{\@secondoftwo}%
\providecommand \href [0]{\begingroup \@sanitize@url \@href}%
\providecommand \@href[1]{\@@startlink{#1}\@@href}%
\providecommand \@@href[1]{\endgroup#1\@@endlink}%
\providecommand \@sanitize@url [0]{\catcode `\\12\catcode `\$12\catcode `\&12\catcode `\#12\catcode `\^12\catcode `\_12\catcode `\%12\relax}%
\providecommand \@@startlink[1]{}%
\providecommand \@@endlink[0]{}%
\providecommand \url  [0]{\begingroup\@sanitize@url \@url }%
\providecommand \@url [1]{\endgroup\@href {#1}{\urlprefix }}%
\providecommand \urlprefix  [0]{URL }%
\providecommand \Eprint [0]{\href }%
\providecommand \doibase [0]{https://doi.org/}%
\providecommand \selectlanguage [0]{\@gobble}%
\providecommand \bibinfo  [0]{\@secondoftwo}%
\providecommand \bibfield  [0]{\@secondoftwo}%
\providecommand \translation [1]{[#1]}%
\providecommand \BibitemOpen [0]{}%
\providecommand \bibitemStop [0]{}%
\providecommand \bibitemNoStop [0]{.\EOS\space}%
\providecommand \EOS [0]{\spacefactor3000\relax}%
\providecommand \BibitemShut  [1]{\csname bibitem#1\endcsname}%
\let\auto@bib@innerbib\@empty
%</preamble>
\bibitem [{\citenamefont {Hasan}\ and\ \citenamefont {Kane}(2010)}]{HasanRMP10}%
  \BibitemOpen
  \bibfield  {author} {\bibinfo {author} {\bibfnamefont {M.~Z.}\ \bibnamefont {Hasan}}\ and\ \bibinfo {author} {\bibfnamefont {C.~L.}\ \bibnamefont {Kane}},\ }\bibfield  {title} {\bibinfo {title} {Colloquium: Topological insulators},\ }\href {https://doi.org/10.1103/RevModPhys.82.3045} {\bibfield  {journal} {\bibinfo  {journal} {Rev. Mod. Phys.}\ }\textbf {\bibinfo {volume} {82}},\ \bibinfo {pages} {3045} (\bibinfo {year} {2010})}\BibitemShut {NoStop}%
\bibitem [{\citenamefont {Qi}\ and\ \citenamefont {Zhang}(2011)}]{QiRMP11}%
  \BibitemOpen
  \bibfield  {author} {\bibinfo {author} {\bibfnamefont {X.-L.}\ \bibnamefont {Qi}}\ and\ \bibinfo {author} {\bibfnamefont {S.-C.}\ \bibnamefont {Zhang}},\ }\bibfield  {title} {\bibinfo {title} {Topological insulators and superconductors},\ }\href {https://doi.org/10.1103/RevModPhys.83.1057} {\bibfield  {journal} {\bibinfo  {journal} {Rev. Mod. Phys.}\ }\textbf {\bibinfo {volume} {83}},\ \bibinfo {pages} {1057} (\bibinfo {year} {2011})}\BibitemShut {NoStop}%
\bibitem [{\citenamefont {Zhang}\ \emph {et~al.}(2009)\citenamefont {Zhang}, \citenamefont {Liu}, \citenamefont {Qi}, \citenamefont {Dai}, \citenamefont {Fang},\ and\ \citenamefont {Zhang}}]{ZhangNaturePhys09}%
  \BibitemOpen
  \bibfield  {author} {\bibinfo {author} {\bibfnamefont {H.}~\bibnamefont {Zhang}}, \bibinfo {author} {\bibfnamefont {C.-X.}\ \bibnamefont {Liu}}, \bibinfo {author} {\bibfnamefont {X.-L.}\ \bibnamefont {Qi}}, \bibinfo {author} {\bibfnamefont {X.}~\bibnamefont {Dai}}, \bibinfo {author} {\bibfnamefont {Z.}~\bibnamefont {Fang}},\ and\ \bibinfo {author} {\bibfnamefont {S.-C.}\ \bibnamefont {Zhang}},\ }\bibfield  {title} {\bibinfo {title} {Topological insulators in {Bi$_2$Se$_3$}, {Bi$_2$Te$_3$} and {Sb$_2$Te$_3$} with a single {Dirac} cone on the surface},\ }\href {https://doi.org/10.1038/nphys1270} {\bibfield  {journal} {\bibinfo  {journal} {Nature Phys.}\ }\textbf {\bibinfo {volume} {5}},\ \bibinfo {pages} {438} (\bibinfo {year} {2009})}\BibitemShut {NoStop}%
\bibitem [{\citenamefont {Hsieh}\ \emph {et~al.}(2009{\natexlab{a}})\citenamefont {Hsieh}, \citenamefont {Xia}, \citenamefont {Qian}, \citenamefont {Wray}, \citenamefont {Dil}, \citenamefont {Meier}, \citenamefont {Osterwalder}, \citenamefont {Patthey}, \citenamefont {Checkelsky}, \citenamefont {Ong} \emph {et~al.}}]{HsiehNature09}%
  \BibitemOpen
  \bibfield  {author} {\bibinfo {author} {\bibfnamefont {D.}~\bibnamefont {Hsieh}}, \bibinfo {author} {\bibfnamefont {Y.}~\bibnamefont {Xia}}, \bibinfo {author} {\bibfnamefont {D.}~\bibnamefont {Qian}}, \bibinfo {author} {\bibfnamefont {L.}~\bibnamefont {Wray}}, \bibinfo {author} {\bibfnamefont {J.}~\bibnamefont {Dil}}, \bibinfo {author} {\bibfnamefont {F.}~\bibnamefont {Meier}}, \bibinfo {author} {\bibfnamefont {J.}~\bibnamefont {Osterwalder}}, \bibinfo {author} {\bibfnamefont {L.}~\bibnamefont {Patthey}}, \bibinfo {author} {\bibfnamefont {J.}~\bibnamefont {Checkelsky}}, \bibinfo {author} {\bibfnamefont {N.~P.}\ \bibnamefont {Ong}}, \emph {et~al.},\ }\bibfield  {title} {\bibinfo {title} {A tunable topological insulator in the spin helical {Dirac} transport regime},\ }\href {https://doi.org/10.1038/nature08234} {\bibfield  {journal} {\bibinfo  {journal} {Nature}\ }\textbf {\bibinfo {volume} {460}},\ \bibinfo {pages} {1101} (\bibinfo {year} {2009}{\natexlab{a}})}\BibitemShut {NoStop}%
\bibitem [{\citenamefont {Chen}\ \emph {et~al.}(2009)\citenamefont {Chen}, \citenamefont {Analytis}, \citenamefont {Chu}, \citenamefont {Liu}, \citenamefont {Mo}, \citenamefont {Qi}, \citenamefont {Zhang}, \citenamefont {Lu}, \citenamefont {Dai}, \citenamefont {Fang}, \citenamefont {Zhang}, \citenamefont {Fisher}, \citenamefont {Hussain},\ and\ \citenamefont {Shen}}]{ChenScience09}%
  \BibitemOpen
  \bibfield  {author} {\bibinfo {author} {\bibfnamefont {Y.~L.}\ \bibnamefont {Chen}}, \bibinfo {author} {\bibfnamefont {J.~G.}\ \bibnamefont {Analytis}}, \bibinfo {author} {\bibfnamefont {J.-H.}\ \bibnamefont {Chu}}, \bibinfo {author} {\bibfnamefont {Z.~K.}\ \bibnamefont {Liu}}, \bibinfo {author} {\bibfnamefont {S.-K.}\ \bibnamefont {Mo}}, \bibinfo {author} {\bibfnamefont {X.~L.}\ \bibnamefont {Qi}}, \bibinfo {author} {\bibfnamefont {H.~J.}\ \bibnamefont {Zhang}}, \bibinfo {author} {\bibfnamefont {D.~H.}\ \bibnamefont {Lu}}, \bibinfo {author} {\bibfnamefont {X.}~\bibnamefont {Dai}}, \bibinfo {author} {\bibfnamefont {Z.}~\bibnamefont {Fang}}, \bibinfo {author} {\bibfnamefont {S.~C.}\ \bibnamefont {Zhang}}, \bibinfo {author} {\bibfnamefont {I.~R.}\ \bibnamefont {Fisher}}, \bibinfo {author} {\bibfnamefont {Z.}~\bibnamefont {Hussain}},\ and\ \bibinfo {author} {\bibfnamefont {Z.-X.}\ \bibnamefont {Shen}},\ }\bibfield  {title} {\bibinfo {title} {Experimental realization of a three-dimensional topological
  insulator, {Bi$_2$Te$_3$}},\ }\href {https://doi.org/10.1126/science.1173034} {\bibfield  {journal} {\bibinfo  {journal} {Science}\ }\textbf {\bibinfo {volume} {325}},\ \bibinfo {pages} {178} (\bibinfo {year} {2009})}\BibitemShut {NoStop}%
\bibitem [{\citenamefont {Hsieh}\ \emph {et~al.}(2009{\natexlab{b}})\citenamefont {Hsieh}, \citenamefont {Xia}, \citenamefont {Qian}, \citenamefont {Wray}, \citenamefont {Meier}, \citenamefont {Dil}, \citenamefont {Osterwalder}, \citenamefont {Patthey}, \citenamefont {Fedorov}, \citenamefont {Lin}, \citenamefont {Bansil}, \citenamefont {Grauer}, \citenamefont {Hor}, \citenamefont {Cava},\ and\ \citenamefont {Hasan}}]{HsiehPRL09}%
  \BibitemOpen
  \bibfield  {author} {\bibinfo {author} {\bibfnamefont {D.}~\bibnamefont {Hsieh}}, \bibinfo {author} {\bibfnamefont {Y.}~\bibnamefont {Xia}}, \bibinfo {author} {\bibfnamefont {D.}~\bibnamefont {Qian}}, \bibinfo {author} {\bibfnamefont {L.}~\bibnamefont {Wray}}, \bibinfo {author} {\bibfnamefont {F.}~\bibnamefont {Meier}}, \bibinfo {author} {\bibfnamefont {J.~H.}\ \bibnamefont {Dil}}, \bibinfo {author} {\bibfnamefont {J.}~\bibnamefont {Osterwalder}}, \bibinfo {author} {\bibfnamefont {L.}~\bibnamefont {Patthey}}, \bibinfo {author} {\bibfnamefont {A.~V.}\ \bibnamefont {Fedorov}}, \bibinfo {author} {\bibfnamefont {H.}~\bibnamefont {Lin}}, \bibinfo {author} {\bibfnamefont {A.}~\bibnamefont {Bansil}}, \bibinfo {author} {\bibfnamefont {D.}~\bibnamefont {Grauer}}, \bibinfo {author} {\bibfnamefont {Y.~S.}\ \bibnamefont {Hor}}, \bibinfo {author} {\bibfnamefont {R.~J.}\ \bibnamefont {Cava}},\ and\ \bibinfo {author} {\bibfnamefont {M.~Z.}\ \bibnamefont {Hasan}},\ }\bibfield  {title} {\bibinfo {title} {Observation of
  time-reversal-protected {single-Dirac-cone} topological-insulator states in {${\mathrm{Bi}}_{2}{\mathrm{Te}}_{3}$ and ${\mathrm{Sb}}_{2}{\mathrm{Te}}_{3}$}},\ }\href {https://doi.org/10.1103/PhysRevLett.103.146401} {\bibfield  {journal} {\bibinfo  {journal} {Phys. Rev. Lett.}\ }\textbf {\bibinfo {volume} {103}},\ \bibinfo {pages} {146401} (\bibinfo {year} {2009}{\natexlab{b}})}\BibitemShut {NoStop}%
\bibitem [{\citenamefont {Orlita}\ \emph {et~al.}(2015)\citenamefont {Orlita}, \citenamefont {Piot}, \citenamefont {Martinez}, \citenamefont {Kumar}, \citenamefont {Faugeras}, \citenamefont {Potemski}, \citenamefont {Michel}, \citenamefont {Hankiewicz}, \citenamefont {Brauner}, \citenamefont {Dra\v{s}ar}, \citenamefont {Schreyeck}, \citenamefont {Grauer}, \citenamefont {Brunner}, \citenamefont {Gould}, \citenamefont {Br\"une},\ and\ \citenamefont {Molenkamp}}]{OrlitaPRL15}%
  \BibitemOpen
  \bibfield  {author} {\bibinfo {author} {\bibfnamefont {M.}~\bibnamefont {Orlita}}, \bibinfo {author} {\bibfnamefont {B.~A.}\ \bibnamefont {Piot}}, \bibinfo {author} {\bibfnamefont {G.}~\bibnamefont {Martinez}}, \bibinfo {author} {\bibfnamefont {N.~K.~S.}\ \bibnamefont {Kumar}}, \bibinfo {author} {\bibfnamefont {C.}~\bibnamefont {Faugeras}}, \bibinfo {author} {\bibfnamefont {M.}~\bibnamefont {Potemski}}, \bibinfo {author} {\bibfnamefont {C.}~\bibnamefont {Michel}}, \bibinfo {author} {\bibfnamefont {E.~M.}\ \bibnamefont {Hankiewicz}}, \bibinfo {author} {\bibfnamefont {T.}~\bibnamefont {Brauner}}, \bibinfo {author} {\bibfnamefont {C.}~\bibnamefont {Dra\v{s}ar}}, \bibinfo {author} {\bibfnamefont {S.}~\bibnamefont {Schreyeck}}, \bibinfo {author} {\bibfnamefont {S.}~\bibnamefont {Grauer}}, \bibinfo {author} {\bibfnamefont {K.}~\bibnamefont {Brunner}}, \bibinfo {author} {\bibfnamefont {C.}~\bibnamefont {Gould}}, \bibinfo {author} {\bibfnamefont {C.}~\bibnamefont {Br\"une}},\ and\ \bibinfo {author} {\bibfnamefont
  {L.~W.}\ \bibnamefont {Molenkamp}},\ }\bibfield  {title} {\bibinfo {title} {Magneto-optics of massive {Dirac} fermions in bulk {Bi$_2$Se$_3$}},\ }\href {https://doi.org/10.1103/PhysRevLett.114.186401} {\bibfield  {journal} {\bibinfo  {journal} {Phys. Rev. Lett.}\ }\textbf {\bibinfo {volume} {114}},\ \bibinfo {pages} {186401} (\bibinfo {year} {2015})}\BibitemShut {NoStop}%
\bibitem [{\citenamefont {Mohelsk\'y}\ \emph {et~al.}(2020)\citenamefont {Mohelsk\'y}, \citenamefont {Dubroka}, \citenamefont {Wyzula}, \citenamefont {Slobodeniuk}, \citenamefont {Martinez}, \citenamefont {Krupko}, \citenamefont {Piot}, \citenamefont {Caha}, \citenamefont {Huml\'{\i}\ifmmode~\check{c}\else \v{c}\fi{}ek}, \citenamefont {Bauer}, \citenamefont {Springholz},\ and\ \citenamefont {Orlita}}]{MohelskyPRB20}%
  \BibitemOpen
  \bibfield  {author} {\bibinfo {author} {\bibfnamefont {I.}~\bibnamefont {Mohelsk\'y}}, \bibinfo {author} {\bibfnamefont {A.}~\bibnamefont {Dubroka}}, \bibinfo {author} {\bibfnamefont {J.}~\bibnamefont {Wyzula}}, \bibinfo {author} {\bibfnamefont {A.}~\bibnamefont {Slobodeniuk}}, \bibinfo {author} {\bibfnamefont {G.}~\bibnamefont {Martinez}}, \bibinfo {author} {\bibfnamefont {Y.}~\bibnamefont {Krupko}}, \bibinfo {author} {\bibfnamefont {B.~A.}\ \bibnamefont {Piot}}, \bibinfo {author} {\bibfnamefont {O.}~\bibnamefont {Caha}}, \bibinfo {author} {\bibfnamefont {J.}~\bibnamefont {Huml\'{\i}\ifmmode~\check{c}\else \v{c}\fi{}ek}}, \bibinfo {author} {\bibfnamefont {G.}~\bibnamefont {Bauer}}, \bibinfo {author} {\bibfnamefont {G.}~\bibnamefont {Springholz}},\ and\ \bibinfo {author} {\bibfnamefont {M.}~\bibnamefont {Orlita}},\ }\bibfield  {title} {\bibinfo {title} {Landau level spectroscopy of {Bi$_2$Te$_3$}},\ }\href {https://doi.org/10.1103/PhysRevB.102.085201} {\bibfield  {journal} {\bibinfo  {journal} {Phys. Rev.
  B}\ }\textbf {\bibinfo {volume} {102}},\ \bibinfo {pages} {085201} (\bibinfo {year} {2020})}\BibitemShut {NoStop}%
\bibitem [{\citenamefont {{K\"{o}hler}}(1976{\natexlab{a}})}]{Kohlerpssb76}%
  \BibitemOpen
  \bibfield  {author} {\bibinfo {author} {\bibfnamefont {H.}~\bibnamefont {{K\"{o}hler}}},\ }\bibfield  {title} {\bibinfo {title} {Non-parabolic {$E(k)$} relation of the lowest conduction band in {Bi$_2$Te$_3$}},\ }\href {https://doi.org/10.1002/pssb.2220730107} {\bibfield  {journal} {\bibinfo  {journal} {phys. stat. sol. (b)}\ }\textbf {\bibinfo {volume} {73}},\ \bibinfo {pages} {95} (\bibinfo {year} {1976}{\natexlab{a}})}\BibitemShut {NoStop}%
\bibitem [{\citenamefont {{K\"{o}hler}}(1976{\natexlab{b}})}]{Kohlerpssb76II}%
  \BibitemOpen
  \bibfield  {author} {\bibinfo {author} {\bibfnamefont {H.}~\bibnamefont {{K\"{o}hler}}},\ }\bibfield  {title} {\bibinfo {title} {Non-parabolicity of the highest valence band of {Bi$_2$Te$_3$} from {Shubnikov-de Haas} effect},\ }\href {https://doi.org/10.1002/pssb.2220740218} {\bibfield  {journal} {\bibinfo  {journal} {phys. stat. sol. (b)}\ }\textbf {\bibinfo {volume} {74}},\ \bibinfo {pages} {591} (\bibinfo {year} {1976}{\natexlab{b}})}\BibitemShut {NoStop}%
\bibitem [{\citenamefont {{K\"{o}hler}}(1976{\natexlab{c}})}]{Kohlerpssb76III}%
  \BibitemOpen
  \bibfield  {author} {\bibinfo {author} {\bibfnamefont {H.}~\bibnamefont {{K\"{o}hler}}},\ }\bibfield  {title} {\bibinfo {title} {Anisotropic $g$-factor of the conduction electrons in {Bi$_2$Te$_3$}},\ }\href {https://doi.org/10.1002/pssb.2220750112} {\bibfield  {journal} {\bibinfo  {journal} {phys. stat. sol. (b)}\ }\textbf {\bibinfo {volume} {75}},\ \bibinfo {pages} {127} (\bibinfo {year} {1976}{\natexlab{c}})}\BibitemShut {NoStop}%
\bibitem [{\citenamefont {Schwartz}\ \emph {et~al.}(1967)\citenamefont {Schwartz}, \citenamefont {{Bj\"{o}rck}},\ and\ \citenamefont {Beckman}}]{SchwartzSSC67}%
  \BibitemOpen
  \bibfield  {author} {\bibinfo {author} {\bibfnamefont {H.}~\bibnamefont {Schwartz}}, \bibinfo {author} {\bibfnamefont {G.}~\bibnamefont {{Bj\"{o}rck}}},\ and\ \bibinfo {author} {\bibfnamefont {O.}~\bibnamefont {Beckman}},\ }\bibfield  {title} {\bibinfo {title} {De {Haas} - van {Alphen} susceptibility measurements on $p$-type {Sb$_2$Te$_3$}},\ }\href {https://doi.org/https://doi.org/10.1016/0038-1098(67)90326-2} {\bibfield  {journal} {\bibinfo  {journal} {Solid State Commun.}\ }\textbf {\bibinfo {volume} {5}},\ \bibinfo {pages} {905} (\bibinfo {year} {1967})}\BibitemShut {NoStop}%
\bibitem [{\citenamefont {{K\"{o}hler}}\ and\ \citenamefont {Freudenberger}(1977)}]{Kohlerpss77}%
  \BibitemOpen
  \bibfield  {author} {\bibinfo {author} {\bibfnamefont {H.}~\bibnamefont {{K\"{o}hler}}}\ and\ \bibinfo {author} {\bibfnamefont {A.}~\bibnamefont {Freudenberger}},\ }\bibfield  {title} {\bibinfo {title} {Investigation of the highest valence band in {(Bi$_{1-x}$Sb$_x$)$_2$Te$_3$} crystals},\ }\href {https://doi.org/https://doi.org/10.1002/pssb.2220840122} {\bibfield  {journal} {\bibinfo  {journal} {phys. stat. sol. (b)}\ }\textbf {\bibinfo {volume} {84}},\ \bibinfo {pages} {195} (\bibinfo {year} {1977})}\BibitemShut {NoStop}%
\bibitem [{\citenamefont {Kulbachinskii}\ \emph {et~al.}(1995)\citenamefont {Kulbachinskii}, \citenamefont {Dashevskii}, \citenamefont {Inoue}, \citenamefont {Sasaki}, \citenamefont {Negishi}, \citenamefont {Gao}, \citenamefont {Lostak}, \citenamefont {Horak},\ and\ \citenamefont {de~Visser}}]{KulbachinskiiPRB95}%
  \BibitemOpen
  \bibfield  {author} {\bibinfo {author} {\bibfnamefont {V.~A.}\ \bibnamefont {Kulbachinskii}}, \bibinfo {author} {\bibfnamefont {Z.~M.}\ \bibnamefont {Dashevskii}}, \bibinfo {author} {\bibfnamefont {M.}~\bibnamefont {Inoue}}, \bibinfo {author} {\bibfnamefont {M.}~\bibnamefont {Sasaki}}, \bibinfo {author} {\bibfnamefont {H.}~\bibnamefont {Negishi}}, \bibinfo {author} {\bibfnamefont {W.~X.}\ \bibnamefont {Gao}}, \bibinfo {author} {\bibfnamefont {P.}~\bibnamefont {Lostak}}, \bibinfo {author} {\bibfnamefont {J.}~\bibnamefont {Horak}},\ and\ \bibinfo {author} {\bibfnamefont {A.}~\bibnamefont {de~Visser}},\ }\bibfield  {title} {\bibinfo {title} {Valence-band changes in {${\mathrm{Sb}}_{2\mathrm{\ensuremath{-}}\mathit{x}}$${\mathrm{In}}_{\mathit{x}}$${\mathrm{Te}}_{3}$} and {${\mathrm{Sb}}_{2}$${\mathrm{Te}}_{3\mathrm{\ensuremath{-}}\mathit{y}}$${\mathrm{Se}}_{\mathit{y}}$} by transport and {Shubnikov--de Haas} effect measurements},\ }\href {https://doi.org/10.1103/PhysRevB.52.10915} {\bibfield  {journal} {\bibinfo
   {journal} {Phys. Rev. B}\ }\textbf {\bibinfo {volume} {52}},\ \bibinfo {pages} {10915} (\bibinfo {year} {1995})}\BibitemShut {NoStop}%
\bibitem [{\citenamefont {Zhao}\ \emph {et~al.}(2019)\citenamefont {Zhao}, \citenamefont {Cortie}, \citenamefont {Chen}, \citenamefont {Li}, \citenamefont {Yue},\ and\ \citenamefont {Wang}}]{ZhaoPRB19}%
  \BibitemOpen
  \bibfield  {author} {\bibinfo {author} {\bibfnamefont {W.}~\bibnamefont {Zhao}}, \bibinfo {author} {\bibfnamefont {D.}~\bibnamefont {Cortie}}, \bibinfo {author} {\bibfnamefont {L.}~\bibnamefont {Chen}}, \bibinfo {author} {\bibfnamefont {Z.}~\bibnamefont {Li}}, \bibinfo {author} {\bibfnamefont {Z.}~\bibnamefont {Yue}},\ and\ \bibinfo {author} {\bibfnamefont {X.}~\bibnamefont {Wang}},\ }\bibfield  {title} {\bibinfo {title} {Quantum oscillations in iron-doped single crystals of the topological insulator $\mathrm{S}{\mathrm{b}}_{2}\mathrm{T}{\mathrm{e}}_{3}$},\ }\href {https://doi.org/10.1103/PhysRevB.99.165133} {\bibfield  {journal} {\bibinfo  {journal} {Phys. Rev. B}\ }\textbf {\bibinfo {volume} {99}},\ \bibinfo {pages} {165133} (\bibinfo {year} {2019})}\BibitemShut {NoStop}%
\bibitem [{\citenamefont {Kulbachinskii}\ \emph {et~al.}(2021)\citenamefont {Kulbachinskii}, \citenamefont {Zinoviev}, \citenamefont {Kytin}, \citenamefont {Mikhailov},\ and\ \citenamefont {Ismailov}}]{KulbachinskiiMTP21}%
  \BibitemOpen
  \bibfield  {author} {\bibinfo {author} {\bibfnamefont {V.}~\bibnamefont {Kulbachinskii}}, \bibinfo {author} {\bibfnamefont {D.}~\bibnamefont {Zinoviev}}, \bibinfo {author} {\bibfnamefont {V.}~\bibnamefont {Kytin}}, \bibinfo {author} {\bibfnamefont {M.}~\bibnamefont {Mikhailov}},\ and\ \bibinfo {author} {\bibfnamefont {Z.}~\bibnamefont {Ismailov}},\ }\bibfield  {title} {\bibinfo {title} {Thermoelectical properties and {Shubnikov - de Haas} effect in single crystals {Sb$_{2-x}$Cu$_x$Te$_3$}},\ }\href {https://doi.org/https://doi.org/10.1016/j.matpr.2020.01.514} {\bibfield  {journal} {\bibinfo  {journal} {Materials Today: Proceedings}\ }\textbf {\bibinfo {volume} {44}},\ \bibinfo {pages} {3439} (\bibinfo {year} {2021})},\ \bibinfo {note} {{17$^{\mathrm{th}}$} European Thermoelectric Conference}\BibitemShut {NoStop}%
\bibitem [{\citenamefont {{von Middendorff}}\ \emph {et~al.}(1973)\citenamefont {{von Middendorff}}, \citenamefont {Dietrich},\ and\ \citenamefont {Landwehr}}]{MiddendorffSSC73}%
  \BibitemOpen
  \bibfield  {author} {\bibinfo {author} {\bibfnamefont {A.}~\bibnamefont {{von Middendorff}}}, \bibinfo {author} {\bibfnamefont {K.}~\bibnamefont {Dietrich}},\ and\ \bibinfo {author} {\bibfnamefont {G.}~\bibnamefont {Landwehr}},\ }\bibfield  {title} {\bibinfo {title} {{Shubnikov-de Haas} effect in $p$-type {Sb$_2$Te$_3$}},\ }\href {https://doi.org/https://doi.org/10.1016/0038-1098(73)90472-9} {\bibfield  {journal} {\bibinfo  {journal} {Solid State Commun.}\ }\textbf {\bibinfo {volume} {13}},\ \bibinfo {pages} {443} (\bibinfo {year} {1973})}\BibitemShut {NoStop}%
\bibitem [{\citenamefont {Simon}\ and\ \citenamefont {Eichler}(1981)}]{Simonpssb81}%
  \BibitemOpen
  \bibfield  {author} {\bibinfo {author} {\bibfnamefont {G.}~\bibnamefont {Simon}}\ and\ \bibinfo {author} {\bibfnamefont {W.}~\bibnamefont {Eichler}},\ }\bibfield  {title} {\bibinfo {title} {Investigations on a two-valence band model for {Sb$_2$Te$_3$}},\ }\href {https://doi.org/https://doi.org/10.1002/pssb.2221070120} {\bibfield  {journal} {\bibinfo  {journal} {phys. stat. sol. (b)}\ }\textbf {\bibinfo {volume} {107}},\ \bibinfo {pages} {201} (\bibinfo {year} {1981})}\BibitemShut {NoStop}%
\bibitem [{\citenamefont {Wang}\ and\ \citenamefont {Cagin}(2007)}]{WangPRB07}%
  \BibitemOpen
  \bibfield  {author} {\bibinfo {author} {\bibfnamefont {G.}~\bibnamefont {Wang}}\ and\ \bibinfo {author} {\bibfnamefont {T.}~\bibnamefont {Cagin}},\ }\bibfield  {title} {\bibinfo {title} {Electronic structure of the thermoelectric materials {${\mathrm{Bi}}_{2}{\mathrm{Te}}_{3}$} and {${\mathrm{Sb}}_{2}{\mathrm{Te}}_{3}$} from first-principles calculations},\ }\href {https://doi.org/10.1103/PhysRevB.76.075201} {\bibfield  {journal} {\bibinfo  {journal} {Phys. Rev. B}\ }\textbf {\bibinfo {volume} {76}},\ \bibinfo {pages} {075201} (\bibinfo {year} {2007})}\BibitemShut {NoStop}%
\bibitem [{\citenamefont {Yavorsky}\ \emph {et~al.}(2011)\citenamefont {Yavorsky}, \citenamefont {Hinsche}, \citenamefont {Mertig},\ and\ \citenamefont {Zahn}}]{YavorskyPRB11}%
  \BibitemOpen
  \bibfield  {author} {\bibinfo {author} {\bibfnamefont {B.~Y.}\ \bibnamefont {Yavorsky}}, \bibinfo {author} {\bibfnamefont {N.~F.}\ \bibnamefont {Hinsche}}, \bibinfo {author} {\bibfnamefont {I.}~\bibnamefont {Mertig}},\ and\ \bibinfo {author} {\bibfnamefont {P.}~\bibnamefont {Zahn}},\ }\bibfield  {title} {\bibinfo {title} {Electronic structure and transport anisotropy of {Bi$_2$Te$_3$ and Sb$_2$Te$_3$}},\ }\href {https://doi.org/10.1103/PhysRevB.84.165208} {\bibfield  {journal} {\bibinfo  {journal} {Phys. Rev. B}\ }\textbf {\bibinfo {volume} {84}},\ \bibinfo {pages} {165208} (\bibinfo {year} {2011})}\BibitemShut {NoStop}%
\bibitem [{\citenamefont {Aguilera}\ \emph {et~al.}(2013{\natexlab{a}})\citenamefont {Aguilera}, \citenamefont {Friedrich}, \citenamefont {Bihlmayer},\ and\ \citenamefont {Bl\"ugel}}]{AguileraPRB13}%
  \BibitemOpen
  \bibfield  {author} {\bibinfo {author} {\bibfnamefont {I.}~\bibnamefont {Aguilera}}, \bibinfo {author} {\bibfnamefont {C.}~\bibnamefont {Friedrich}}, \bibinfo {author} {\bibfnamefont {G.}~\bibnamefont {Bihlmayer}},\ and\ \bibinfo {author} {\bibfnamefont {S.}~\bibnamefont {Bl\"ugel}},\ }\bibfield  {title} {\bibinfo {title} {{$GW$} study of topological insulators {Bi${}_{2}$Se${}_{3}$}, {Bi${}_{2}$Te${}_{3}$}, and {Sb${}_{2}$Te${}_{3}$}: Beyond the perturbative one-shot approach},\ }\href {https://doi.org/10.1103/PhysRevB.88.045206} {\bibfield  {journal} {\bibinfo  {journal} {Phys. Rev. B}\ }\textbf {\bibinfo {volume} {88}},\ \bibinfo {pages} {045206} (\bibinfo {year} {2013}{\natexlab{a}})}\BibitemShut {NoStop}%
\bibitem [{\citenamefont {Nechaev}\ \emph {et~al.}(2015)\citenamefont {Nechaev}, \citenamefont {Aguilera}, \citenamefont {De~Renzi}, \citenamefont {di~Bona}, \citenamefont {Lodi~Rizzini}, \citenamefont {Mio}, \citenamefont {Nicotra}, \citenamefont {Politano}, \citenamefont {Scalese}, \citenamefont {Aliev}, \citenamefont {Babanly}, \citenamefont {Friedrich}, \citenamefont {Bl\"ugel},\ and\ \citenamefont {Chulkov}}]{NechaevPRB15}%
  \BibitemOpen
  \bibfield  {author} {\bibinfo {author} {\bibfnamefont {I.~A.}\ \bibnamefont {Nechaev}}, \bibinfo {author} {\bibfnamefont {I.}~\bibnamefont {Aguilera}}, \bibinfo {author} {\bibfnamefont {V.}~\bibnamefont {De~Renzi}}, \bibinfo {author} {\bibfnamefont {A.}~\bibnamefont {di~Bona}}, \bibinfo {author} {\bibfnamefont {A.}~\bibnamefont {Lodi~Rizzini}}, \bibinfo {author} {\bibfnamefont {A.~M.}\ \bibnamefont {Mio}}, \bibinfo {author} {\bibfnamefont {G.}~\bibnamefont {Nicotra}}, \bibinfo {author} {\bibfnamefont {A.}~\bibnamefont {Politano}}, \bibinfo {author} {\bibfnamefont {S.}~\bibnamefont {Scalese}}, \bibinfo {author} {\bibfnamefont {Z.~S.}\ \bibnamefont {Aliev}}, \bibinfo {author} {\bibfnamefont {M.~B.}\ \bibnamefont {Babanly}}, \bibinfo {author} {\bibfnamefont {C.}~\bibnamefont {Friedrich}}, \bibinfo {author} {\bibfnamefont {S.}~\bibnamefont {Bl\"ugel}},\ and\ \bibinfo {author} {\bibfnamefont {E.~V.}\ \bibnamefont {Chulkov}},\ }\bibfield  {title} {\bibinfo {title} {Quasiparticle spectrum and plasmonic excitations
  in the topological insulator {${\mathrm{Sb}}_{2}{\mathrm{Te}}_{3}$}},\ }\href {https://doi.org/10.1103/PhysRevB.91.245123} {\bibfield  {journal} {\bibinfo  {journal} {Phys. Rev. B}\ }\textbf {\bibinfo {volume} {91}},\ \bibinfo {pages} {245123} (\bibinfo {year} {2015})}\BibitemShut {NoStop}%
\bibitem [{\citenamefont {Aguilera}\ \emph {et~al.}(2019)\citenamefont {Aguilera}, \citenamefont {Friedrich},\ and\ \citenamefont {Bl\"ugel}}]{AguileraPRB19}%
  \BibitemOpen
  \bibfield  {author} {\bibinfo {author} {\bibfnamefont {I.}~\bibnamefont {Aguilera}}, \bibinfo {author} {\bibfnamefont {C.}~\bibnamefont {Friedrich}},\ and\ \bibinfo {author} {\bibfnamefont {S.}~\bibnamefont {Bl\"ugel}},\ }\bibfield  {title} {\bibinfo {title} {Many-body corrected tight-binding hamiltonians for an accurate quasiparticle description of topological insulators of the {${\mathrm{Bi}}_{2}{\mathrm{Se}}_{3}$} family},\ }\href {https://doi.org/10.1103/PhysRevB.100.155147} {\bibfield  {journal} {\bibinfo  {journal} {Phys. Rev. B}\ }\textbf {\bibinfo {volume} {100}},\ \bibinfo {pages} {155147} (\bibinfo {year} {2019})}\BibitemShut {NoStop}%
\bibitem [{\citenamefont {Anderson}\ and\ \citenamefont {Krause}(1974)}]{AndersonACSB74}%
  \BibitemOpen
  \bibfield  {author} {\bibinfo {author} {\bibfnamefont {T.~L.}\ \bibnamefont {Anderson}}\ and\ \bibinfo {author} {\bibfnamefont {H.~B.}\ \bibnamefont {Krause}},\ }\bibfield  {title} {\bibinfo {title} {{Refinement of the Sb${\sb 2}$Te${\sb 3}$ and Sb${\sb 2}$Te${\sb 2}$Se structures and their relationship to nonstoichiometric {Sb$_2$Te$_{3-y}$Se$_y$} compounds}},\ }\href {https://doi.org/10.1107/S0567740874004729} {\bibfield  {journal} {\bibinfo  {journal} {Acta Crystallographica Section B}\ }\textbf {\bibinfo {volume} {30}},\ \bibinfo {pages} {1307} (\bibinfo {year} {1974})}\BibitemShut {NoStop}%
\bibitem [{\citenamefont {Bragaglia}\ \emph {et~al.}(2020)\citenamefont {Bragaglia}, \citenamefont {Ramsteiner}, \citenamefont {Schick}, \citenamefont {Boschker}, \citenamefont {Mitzner}, \citenamefont {Calarco},\ and\ \citenamefont {Holldack}}]{BragagliaSR20}%
  \BibitemOpen
  \bibfield  {author} {\bibinfo {author} {\bibfnamefont {V.}~\bibnamefont {Bragaglia}}, \bibinfo {author} {\bibfnamefont {M.}~\bibnamefont {Ramsteiner}}, \bibinfo {author} {\bibfnamefont {D.}~\bibnamefont {Schick}}, \bibinfo {author} {\bibfnamefont {J.~E.}\ \bibnamefont {Boschker}}, \bibinfo {author} {\bibfnamefont {R.}~\bibnamefont {Mitzner}}, \bibinfo {author} {\bibfnamefont {R.}~\bibnamefont {Calarco}},\ and\ \bibinfo {author} {\bibfnamefont {K.}~\bibnamefont {Holldack}},\ }\bibfield  {title} {\bibinfo {title} {Phonon anharmonicities and ultrafast dynamics in epitaxial {Sb$_2$Te$_3$}},\ }\href {https://doi.org/10.1038/s41598-020-69663-y} {\bibfield  {journal} {\bibinfo  {journal} {Sci. Rep.}\ }\textbf {\bibinfo {volume} {10}},\ \bibinfo {pages} {12962} (\bibinfo {year} {2020})}\BibitemShut {NoStop}%
\bibitem [{\citenamefont {Dubroka}\ \emph {et~al.}(2017)\citenamefont {Dubroka}, \citenamefont {Caha}, \citenamefont {Hron\ifmmode~\check{c}\else \v{c}\fi{}ek}, \citenamefont {Fri\ifmmode~\check{s}\else \v{s}\fi{}}, \citenamefont {Orlita}, \citenamefont {Hol\'y}, \citenamefont {Steiner}, \citenamefont {Bauer}, \citenamefont {Springholz},\ and\ \citenamefont {Huml\'{\i}\ifmmode~\check{c}\else \v{c}\fi{}ek}}]{DubrokaPRB17}%
  \BibitemOpen
  \bibfield  {author} {\bibinfo {author} {\bibfnamefont {A.}~\bibnamefont {Dubroka}}, \bibinfo {author} {\bibfnamefont {O.}~\bibnamefont {Caha}}, \bibinfo {author} {\bibfnamefont {M.}~\bibnamefont {Hron\ifmmode~\check{c}\else \v{c}\fi{}ek}}, \bibinfo {author} {\bibfnamefont {P.}~\bibnamefont {Fri\ifmmode~\check{s}\else \v{s}\fi{}}}, \bibinfo {author} {\bibfnamefont {M.}~\bibnamefont {Orlita}}, \bibinfo {author} {\bibfnamefont {V.}~\bibnamefont {Hol\'y}}, \bibinfo {author} {\bibfnamefont {H.}~\bibnamefont {Steiner}}, \bibinfo {author} {\bibfnamefont {G.}~\bibnamefont {Bauer}}, \bibinfo {author} {\bibfnamefont {G.}~\bibnamefont {Springholz}},\ and\ \bibinfo {author} {\bibfnamefont {J.}~\bibnamefont {Huml\'{\i}\ifmmode~\check{c}\else \v{c}\fi{}ek}},\ }\bibfield  {title} {\bibinfo {title} {Interband absorption edge in the topological insulators {Bi$_2$(Te$_{1-x}$Se$_x$)$_3$ }},\ }\href {https://doi.org/10.1103/PhysRevB.96.235202} {\bibfield  {journal} {\bibinfo  {journal} {Phys. Rev. B}\ }\textbf {\bibinfo
  {volume} {96}},\ \bibinfo {pages} {235202} (\bibinfo {year} {2017})}\BibitemShut {NoStop}%
\bibitem [{fle()}]{fleur}%
  \BibitemOpen
  \href@noop {} {}\bibinfo {howpublished} {\url{www.flapw.de}}\BibitemShut {NoStop}%
\bibitem [{\citenamefont {Friedrich}\ \emph {et~al.}(2010)\citenamefont {Friedrich}, \citenamefont {Bl\"ugel},\ and\ \citenamefont {Schindlmayr}}]{FriedrichPRB10}%
  \BibitemOpen
  \bibfield  {author} {\bibinfo {author} {\bibfnamefont {C.}~\bibnamefont {Friedrich}}, \bibinfo {author} {\bibfnamefont {S.}~\bibnamefont {Bl\"ugel}},\ and\ \bibinfo {author} {\bibfnamefont {A.}~\bibnamefont {Schindlmayr}},\ }\bibfield  {title} {\bibinfo {title} {Efficient implementation of the {$GW$} approximation within the all-electron flapw method},\ }\href {https://doi.org/10.1103/PhysRevB.81.125102} {\bibfield  {journal} {\bibinfo  {journal} {Phys. Rev. B}\ }\textbf {\bibinfo {volume} {81}},\ \bibinfo {pages} {125102} (\bibinfo {year} {2010})}\BibitemShut {NoStop}%
\bibitem [{\citenamefont {Perdew}\ \emph {et~al.}(1996)\citenamefont {Perdew}, \citenamefont {Burke},\ and\ \citenamefont {Ernzerhof}}]{PerdewPRL96}%
  \BibitemOpen
  \bibfield  {author} {\bibinfo {author} {\bibfnamefont {J.~P.}\ \bibnamefont {Perdew}}, \bibinfo {author} {\bibfnamefont {K.}~\bibnamefont {Burke}},\ and\ \bibinfo {author} {\bibfnamefont {M.}~\bibnamefont {Ernzerhof}},\ }\bibfield  {title} {\bibinfo {title} {Generalized gradient approximation made simple},\ }\href {https://doi.org/10.1103/PhysRevLett.77.3865} {\bibfield  {journal} {\bibinfo  {journal} {Phys. Rev. Lett.}\ }\textbf {\bibinfo {volume} {77}},\ \bibinfo {pages} {3865} (\bibinfo {year} {1996})}\BibitemShut {NoStop}%
\bibitem [{\citenamefont {Aguilera}\ \emph {et~al.}(2013{\natexlab{b}})\citenamefont {Aguilera}, \citenamefont {Friedrich},\ and\ \citenamefont {Bl\"ugel}}]{AguileraPRB13II}%
  \BibitemOpen
  \bibfield  {author} {\bibinfo {author} {\bibfnamefont {I.}~\bibnamefont {Aguilera}}, \bibinfo {author} {\bibfnamefont {C.}~\bibnamefont {Friedrich}},\ and\ \bibinfo {author} {\bibfnamefont {S.}~\bibnamefont {Bl\"ugel}},\ }\bibfield  {title} {\bibinfo {title} {Spin-orbit coupling in quasiparticle studies of topological insulators},\ }\href {https://doi.org/10.1103/PhysRevB.88.165136} {\bibfield  {journal} {\bibinfo  {journal} {Phys. Rev. B}\ }\textbf {\bibinfo {volume} {88}},\ \bibinfo {pages} {165136} (\bibinfo {year} {2013}{\natexlab{b}})}\BibitemShut {NoStop}%
\bibitem [{\citenamefont {Sakuma}\ \emph {et~al.}(2011)\citenamefont {Sakuma}, \citenamefont {Friedrich}, \citenamefont {Miyake}, \citenamefont {Bl\"ugel},\ and\ \citenamefont {Aryasetiawan}}]{SakumaPRB11}%
  \BibitemOpen
  \bibfield  {author} {\bibinfo {author} {\bibfnamefont {R.}~\bibnamefont {Sakuma}}, \bibinfo {author} {\bibfnamefont {C.}~\bibnamefont {Friedrich}}, \bibinfo {author} {\bibfnamefont {T.}~\bibnamefont {Miyake}}, \bibinfo {author} {\bibfnamefont {S.}~\bibnamefont {Bl\"ugel}},\ and\ \bibinfo {author} {\bibfnamefont {F.}~\bibnamefont {Aryasetiawan}},\ }\bibfield  {title} {\bibinfo {title} {{$GW$} calculations including spin-orbit coupling: {Application} to {Hg} chalcogenides},\ }\href {https://doi.org/10.1103/PhysRevB.84.085144} {\bibfield  {journal} {\bibinfo  {journal} {Phys. Rev. B}\ }\textbf {\bibinfo {volume} {84}},\ \bibinfo {pages} {085144} (\bibinfo {year} {2011})}\BibitemShut {NoStop}%
\bibitem [{\citenamefont {Marzari}\ \emph {et~al.}(2012)\citenamefont {Marzari}, \citenamefont {Mostofi}, \citenamefont {Yates}, \citenamefont {Souza},\ and\ \citenamefont {Vanderbilt}}]{MarzariRMP12}%
  \BibitemOpen
  \bibfield  {author} {\bibinfo {author} {\bibfnamefont {N.}~\bibnamefont {Marzari}}, \bibinfo {author} {\bibfnamefont {A.~A.}\ \bibnamefont {Mostofi}}, \bibinfo {author} {\bibfnamefont {J.~R.}\ \bibnamefont {Yates}}, \bibinfo {author} {\bibfnamefont {I.}~\bibnamefont {Souza}},\ and\ \bibinfo {author} {\bibfnamefont {D.}~\bibnamefont {Vanderbilt}},\ }\bibfield  {title} {\bibinfo {title} {Maximally localized {Wannier} functions: Theory and applications},\ }\href {https://doi.org/10.1103/RevModPhys.84.1419} {\bibfield  {journal} {\bibinfo  {journal} {Rev. Mod. Phys.}\ }\textbf {\bibinfo {volume} {84}},\ \bibinfo {pages} {1419} (\bibinfo {year} {2012})}\BibitemShut {NoStop}%
\bibitem [{\citenamefont {Mostofi}\ \emph {et~al.}(2008)\citenamefont {Mostofi}, \citenamefont {Yates}, \citenamefont {Lee}, \citenamefont {Souza}, \citenamefont {Vanderbilt},\ and\ \citenamefont {Marzari}}]{MostofiCPC08}%
  \BibitemOpen
  \bibfield  {author} {\bibinfo {author} {\bibfnamefont {A.~A.}\ \bibnamefont {Mostofi}}, \bibinfo {author} {\bibfnamefont {J.~R.}\ \bibnamefont {Yates}}, \bibinfo {author} {\bibfnamefont {Y.-S.}\ \bibnamefont {Lee}}, \bibinfo {author} {\bibfnamefont {I.}~\bibnamefont {Souza}}, \bibinfo {author} {\bibfnamefont {D.}~\bibnamefont {Vanderbilt}},\ and\ \bibinfo {author} {\bibfnamefont {N.}~\bibnamefont {Marzari}},\ }\bibfield  {title} {\bibinfo {title} {wannier90: A tool for obtaining maximally-localised {Wannier} functions},\ }\href {https://doi.org/https://doi.org/10.1016/j.cpc.2007.11.016} {\bibfield  {journal} {\bibinfo  {journal} {Comput. Phys. Commun.}\ }\textbf {\bibinfo {volume} {178}},\ \bibinfo {pages} {685} (\bibinfo {year} {2008})}\BibitemShut {NoStop}%
\bibitem [{\citenamefont {Zhu}\ \emph {et~al.}(2015)\citenamefont {Zhu}, \citenamefont {Ishida}, \citenamefont {Kuroda}, \citenamefont {Sumida}, \citenamefont {Ye}, \citenamefont {Wang}, \citenamefont {Pan}, \citenamefont {Taniguchi}, \citenamefont {Qiao}, \citenamefont {Shin},\ and\ \citenamefont {Kimura}}]{ZhuSR15}%
  \BibitemOpen
  \bibfield  {author} {\bibinfo {author} {\bibfnamefont {S.}~\bibnamefont {Zhu}}, \bibinfo {author} {\bibfnamefont {Y.}~\bibnamefont {Ishida}}, \bibinfo {author} {\bibfnamefont {K.}~\bibnamefont {Kuroda}}, \bibinfo {author} {\bibfnamefont {K.}~\bibnamefont {Sumida}}, \bibinfo {author} {\bibfnamefont {M.}~\bibnamefont {Ye}}, \bibinfo {author} {\bibfnamefont {J.}~\bibnamefont {Wang}}, \bibinfo {author} {\bibfnamefont {H.}~\bibnamefont {Pan}}, \bibinfo {author} {\bibfnamefont {M.}~\bibnamefont {Taniguchi}}, \bibinfo {author} {\bibfnamefont {S.}~\bibnamefont {Qiao}}, \bibinfo {author} {\bibfnamefont {S.}~\bibnamefont {Shin}},\ and\ \bibinfo {author} {\bibfnamefont {A.}~\bibnamefont {Kimura}},\ }\bibfield  {title} {\bibinfo {title} {Ultrafast electron dynamics at the {Dirac} node of the topological insulator {Sb$_2$Te$_3$}},\ }\href {https://doi.org/10.1038/srep13213} {\bibfield  {journal} {\bibinfo  {journal} {Sci. Rep.}\ }\textbf {\bibinfo {volume} {5}},\ \bibinfo {pages} {13213} (\bibinfo {year}
  {2015})}\BibitemShut {NoStop}%
\bibitem [{\citenamefont {Lund}\ \emph {et~al.}(2021)\citenamefont {Lund}, \citenamefont {Volckaert}, \citenamefont {Majchrzak}, \citenamefont {Jones}, \citenamefont {Bianchi}, \citenamefont {Bremholm},\ and\ \citenamefont {Hofmann}}]{LundPCCP21}%
  \BibitemOpen
  \bibfield  {author} {\bibinfo {author} {\bibfnamefont {H.~E.}\ \bibnamefont {Lund}}, \bibinfo {author} {\bibfnamefont {K.}~\bibnamefont {Volckaert}}, \bibinfo {author} {\bibfnamefont {P.}~\bibnamefont {Majchrzak}}, \bibinfo {author} {\bibfnamefont {A.~J.~H.}\ \bibnamefont {Jones}}, \bibinfo {author} {\bibfnamefont {M.}~\bibnamefont {Bianchi}}, \bibinfo {author} {\bibfnamefont {M.}~\bibnamefont {Bremholm}},\ and\ \bibinfo {author} {\bibfnamefont {P.}~\bibnamefont {Hofmann}},\ }\bibfield  {title} {\bibinfo {title} {Bulk band structure of {Sb$_2$Te$_3$} determined by angle-resolved photoemission spectroscopy},\ }\href {https://doi.org/10.1039/D1CP04031F} {\bibfield  {journal} {\bibinfo  {journal} {Phys. Chem. Chem. Phys.}\ }\textbf {\bibinfo {volume} {23}},\ \bibinfo {pages} {26401} (\bibinfo {year} {2021})}\BibitemShut {NoStop}%
\bibitem [{\citenamefont {Locatelli}\ \emph {et~al.}(2022)\citenamefont {Locatelli}, \citenamefont {Kumar}, \citenamefont {Tsipas}, \citenamefont {Dimoulas}, \citenamefont {Longo},\ and\ \citenamefont {Mantovan}}]{LocatelliSR22}%
  \BibitemOpen
  \bibfield  {author} {\bibinfo {author} {\bibfnamefont {L.}~\bibnamefont {Locatelli}}, \bibinfo {author} {\bibfnamefont {A.}~\bibnamefont {Kumar}}, \bibinfo {author} {\bibfnamefont {P.}~\bibnamefont {Tsipas}}, \bibinfo {author} {\bibfnamefont {A.}~\bibnamefont {Dimoulas}}, \bibinfo {author} {\bibfnamefont {E.}~\bibnamefont {Longo}},\ and\ \bibinfo {author} {\bibfnamefont {R.}~\bibnamefont {Mantovan}},\ }\bibfield  {title} {\bibinfo {title} {Magnetotransport and {ARPES} studies of the topological insulators {Sb$_2$Te$_3$} and {Bi$_2$Te$_3$} grown by {MOCVD} on large-area {Si} substrates},\ }\href {https://doi.org/10.1038/s41598-022-07496-7} {\bibfield  {journal} {\bibinfo  {journal} {Sci. Rep.}\ }\textbf {\bibinfo {volume} {12}},\ \bibinfo {pages} {3891} (\bibinfo {year} {2022})}\BibitemShut {NoStop}%
\bibitem [{\citenamefont {Ly}\ and\ \citenamefont {Basko}(2016)}]{LyJPCM16}%
  \BibitemOpen
  \bibfield  {author} {\bibinfo {author} {\bibfnamefont {O.}~\bibnamefont {Ly}}\ and\ \bibinfo {author} {\bibfnamefont {D.~M.}\ \bibnamefont {Basko}},\ }\bibfield  {title} {\bibinfo {title} {Theory of electron spin resonance in bulk topological insulators {Bi$_2$Se$_3$}, {Bi$_2$Te$_3$} and {Sb$_2$Te$_3$}},\ }\href {https://doi.org/10.1088/0953-8984/28/15/155801} {\bibfield  {journal} {\bibinfo  {journal} {J. Phys. Condens. Matter}\ }\textbf {\bibinfo {volume} {28}},\ \bibinfo {pages} {155801} (\bibinfo {year} {2016})}\BibitemShut {NoStop}%
\bibitem [{\citenamefont {Liu}\ \emph {et~al.}(2010)\citenamefont {Liu}, \citenamefont {Qi}, \citenamefont {Zhang}, \citenamefont {Dai}, \citenamefont {Fang},\ and\ \citenamefont {Zhang}}]{LiuPRB10}%
  \BibitemOpen
  \bibfield  {author} {\bibinfo {author} {\bibfnamefont {C.-X.}\ \bibnamefont {Liu}}, \bibinfo {author} {\bibfnamefont {X.-L.}\ \bibnamefont {Qi}}, \bibinfo {author} {\bibfnamefont {H.}~\bibnamefont {Zhang}}, \bibinfo {author} {\bibfnamefont {X.}~\bibnamefont {Dai}}, \bibinfo {author} {\bibfnamefont {Z.}~\bibnamefont {Fang}},\ and\ \bibinfo {author} {\bibfnamefont {S.-C.}\ \bibnamefont {Zhang}},\ }\bibfield  {title} {\bibinfo {title} {Model {Hamiltonian} for topological insulators},\ }\href {https://doi.org/10.1103/PhysRevB.82.045122} {\bibfield  {journal} {\bibinfo  {journal} {Phys. Rev. B}\ }\textbf {\bibinfo {volume} {82}},\ \bibinfo {pages} {045122} (\bibinfo {year} {2010})}\BibitemShut {NoStop}%
\bibitem [{\citenamefont {Assaf}\ \emph {et~al.}(2016)\citenamefont {Assaf}, \citenamefont {Phuphachong}, \citenamefont {Volobuev}, \citenamefont {Inhofer}, \citenamefont {Bauer}, \citenamefont {Springholz}, \citenamefont {de~Vaulchier},\ and\ \citenamefont {Guldner}}]{AssafSR16}%
  \BibitemOpen
  \bibfield  {author} {\bibinfo {author} {\bibfnamefont {B.~A.}\ \bibnamefont {Assaf}}, \bibinfo {author} {\bibfnamefont {T.}~\bibnamefont {Phuphachong}}, \bibinfo {author} {\bibfnamefont {V.~V.}\ \bibnamefont {Volobuev}}, \bibinfo {author} {\bibfnamefont {A.}~\bibnamefont {Inhofer}}, \bibinfo {author} {\bibfnamefont {G.}~\bibnamefont {Bauer}}, \bibinfo {author} {\bibfnamefont {G.}~\bibnamefont {Springholz}}, \bibinfo {author} {\bibfnamefont {L.~A.}\ \bibnamefont {de~Vaulchier}},\ and\ \bibinfo {author} {\bibfnamefont {Y.}~\bibnamefont {Guldner}},\ }\bibfield  {title} {\bibinfo {title} {Massive and massless {Dirac} fermions in {Pb$_{1-x}$Sn$_x$Te} topological crystalline insulator probed by magneto-optical absorption},\ }\href {https://doi.org/10.1038/srep20323} {\bibfield  {journal} {\bibinfo  {journal} {Sci. Rep.}\ }\textbf {\bibinfo {volume} {6}},\ \bibinfo {pages} {20323} (\bibinfo {year} {2016})}\BibitemShut {NoStop}%
\bibitem [{\citenamefont {Tikui\ifmmode~\check{s}\else \v{s}\fi{}is}\ \emph {et~al.}(2021)\citenamefont {Tikui\ifmmode~\check{s}\else \v{s}\fi{}is}, \citenamefont {Wyzula}, \citenamefont {Ohnoutek}, \citenamefont {Cejpek}, \citenamefont {Uhl\'{\i}\ifmmode~\check{r}\else \v{r}\fi{}ov\'a}, \citenamefont {Hakl}, \citenamefont {Faugeras}, \citenamefont {V\'yborn\'y}, \citenamefont {Ishida}, \citenamefont {Veis},\ and\ \citenamefont {Orlita}}]{TikuisisPRB21}%
  \BibitemOpen
  \bibfield  {author} {\bibinfo {author} {\bibfnamefont {K.~K.}\ \bibnamefont {Tikui\ifmmode~\check{s}\else \v{s}\fi{}is}}, \bibinfo {author} {\bibfnamefont {J.}~\bibnamefont {Wyzula}}, \bibinfo {author} {\bibfnamefont {L.}~\bibnamefont {Ohnoutek}}, \bibinfo {author} {\bibfnamefont {P.}~\bibnamefont {Cejpek}}, \bibinfo {author} {\bibfnamefont {K.}~\bibnamefont {Uhl\'{\i}\ifmmode~\check{r}\else \v{r}\fi{}ov\'a}}, \bibinfo {author} {\bibfnamefont {M.}~\bibnamefont {Hakl}}, \bibinfo {author} {\bibfnamefont {C.}~\bibnamefont {Faugeras}}, \bibinfo {author} {\bibfnamefont {K.}~\bibnamefont {V\'yborn\'y}}, \bibinfo {author} {\bibfnamefont {A.}~\bibnamefont {Ishida}}, \bibinfo {author} {\bibfnamefont {M.}~\bibnamefont {Veis}},\ and\ \bibinfo {author} {\bibfnamefont {M.}~\bibnamefont {Orlita}},\ }\bibfield  {title} {\bibinfo {title} {Landau level spectroscopy of the {PbSnSe} topological crystalline insulator},\ }\href {https://doi.org/10.1103/PhysRevB.103.155304} {\bibfield  {journal} {\bibinfo  {journal} {Phys. Rev.
  B}\ }\textbf {\bibinfo {volume} {103}},\ \bibinfo {pages} {155304} (\bibinfo {year} {2021})}\BibitemShut {NoStop}%
\bibitem [{\citenamefont {Burstein}(1954)}]{BursteinPR54}%
  \BibitemOpen
  \bibfield  {author} {\bibinfo {author} {\bibfnamefont {E.}~\bibnamefont {Burstein}},\ }\bibfield  {title} {\bibinfo {title} {Anomalous optical absorption limit in {InSb}},\ }\href {https://doi.org/10.1103/PhysRev.93.632} {\bibfield  {journal} {\bibinfo  {journal} {Phys. Rev.}\ }\textbf {\bibinfo {volume} {93}},\ \bibinfo {pages} {632} (\bibinfo {year} {1954})}\BibitemShut {NoStop}%
\bibitem [{\citenamefont {Moss}(1954)}]{MossPPS54}%
  \BibitemOpen
  \bibfield  {author} {\bibinfo {author} {\bibfnamefont {T.~S.}\ \bibnamefont {Moss}},\ }\bibfield  {title} {\bibinfo {title} {The interpretation of the properties of indium antimonide},\ }\href {https://doi.org/10.1088/0370-1301/67/10/306} {\bibfield  {journal} {\bibinfo  {journal} {Proc. Phys. Soc. B}\ }\textbf {\bibinfo {volume} {67}},\ \bibinfo {pages} {775} (\bibinfo {year} {1954})}\BibitemShut {NoStop}%
\bibitem [{\citenamefont {Cava}\ \emph {et~al.}(2013)\citenamefont {Cava}, \citenamefont {Ji}, \citenamefont {Fuccillo}, \citenamefont {Gibson},\ and\ \citenamefont {Hor}}]{CavaJMCC13}%
  \BibitemOpen
  \bibfield  {author} {\bibinfo {author} {\bibfnamefont {R.~J.}\ \bibnamefont {Cava}}, \bibinfo {author} {\bibfnamefont {H.}~\bibnamefont {Ji}}, \bibinfo {author} {\bibfnamefont {M.~K.}\ \bibnamefont {Fuccillo}}, \bibinfo {author} {\bibfnamefont {Q.~D.}\ \bibnamefont {Gibson}},\ and\ \bibinfo {author} {\bibfnamefont {Y.~S.}\ \bibnamefont {Hor}},\ }\bibfield  {title} {\bibinfo {title} {Crystal structure and chemistry of topological insulators},\ }\href {https://doi.org/10.1039/C3TC30186A} {\bibfield  {journal} {\bibinfo  {journal} {J. Mater. Chem. C}\ }\textbf {\bibinfo {volume} {1}},\ \bibinfo {pages} {3176} (\bibinfo {year} {2013})}\BibitemShut {NoStop}%
\bibitem [{\citenamefont {Horak}\ \emph {et~al.}(1995)\citenamefont {Horak}, \citenamefont {Drasar}, \citenamefont {Novotny}, \citenamefont {Karamazov},\ and\ \citenamefont {Lostak}}]{Horakpssa95}%
  \BibitemOpen
  \bibfield  {author} {\bibinfo {author} {\bibfnamefont {J.}~\bibnamefont {Horak}}, \bibinfo {author} {\bibfnamefont {C.}~\bibnamefont {Drasar}}, \bibinfo {author} {\bibfnamefont {R.}~\bibnamefont {Novotny}}, \bibinfo {author} {\bibfnamefont {S.}~\bibnamefont {Karamazov}},\ and\ \bibinfo {author} {\bibfnamefont {P.}~\bibnamefont {Lostak}},\ }\bibfield  {title} {\bibinfo {title} {Non-stoichiometry of the crystal lattice of antimony telluride},\ }\href {https://doi.org/https://doi.org/10.1002/pssa.2211490205} {\bibfield  {journal} {\bibinfo  {journal} {phys. stat. sol. (a)}\ }\textbf {\bibinfo {volume} {149}},\ \bibinfo {pages} {549} (\bibinfo {year} {1995})}\BibitemShut {NoStop}%
\bibitem [{\citenamefont {Rajput}\ \emph {et~al.}(2023)\citenamefont {Rajput}, \citenamefont {Kumar},\ and\ \citenamefont {Lakhani}}]{RajputCGD23}%
  \BibitemOpen
  \bibfield  {author} {\bibinfo {author} {\bibfnamefont {I.}~\bibnamefont {Rajput}}, \bibinfo {author} {\bibfnamefont {D.}~\bibnamefont {Kumar}},\ and\ \bibinfo {author} {\bibfnamefont {A.}~\bibnamefont {Lakhani}},\ }\bibfield  {title} {\bibinfo {title} {Empirical role of crystalline defects in the transport properties of {Sb$_2$Te$_3$} single crystals},\ }\href {https://doi.org/10.1021/acs.cgd.3c00558} {\bibfield  {journal} {\bibinfo  {journal} {Crystal Growth \& Design}\ }\textbf {\bibinfo {volume} {23}},\ \bibinfo {pages} {6019} (\bibinfo {year} {2023})}\BibitemShut {NoStop}%
\bibitem [{\citenamefont {{K\"ohler}}\ and\ \citenamefont {{W\"ochner}}(1975)}]{Kohlerpssb75}%
  \BibitemOpen
  \bibfield  {author} {\bibinfo {author} {\bibfnamefont {H.}~\bibnamefont {{K\"ohler}}}\ and\ \bibinfo {author} {\bibfnamefont {E.}~\bibnamefont {{W\"ochner}}},\ }\bibfield  {title} {\bibinfo {title} {The $g$-factor of the conduction electrons in {Bi$_2$Se$_3$}},\ }\href {https://doi.org/https://doi.org/10.1002/pssb.2220670229} {\bibfield  {journal} {\bibinfo  {journal} {phys. stat. sol. (b)}\ }\textbf {\bibinfo {volume} {67}},\ \bibinfo {pages} {665} (\bibinfo {year} {1975})}\BibitemShut {NoStop}%
\bibitem [{\citenamefont {Weyrich}\ \emph {et~al.}(2017)\citenamefont {Weyrich}, \citenamefont {Merzenich}, \citenamefont {Kampmeier}, \citenamefont {Batov}, \citenamefont {Mussler}, \citenamefont {Schubert}, \citenamefont {Gr\"utzmacher},\ and\ \citenamefont {Sch\"apers}}]{WeyrichAPL17}%
  \BibitemOpen
  \bibfield  {author} {\bibinfo {author} {\bibfnamefont {C.}~\bibnamefont {Weyrich}}, \bibinfo {author} {\bibfnamefont {T.}~\bibnamefont {Merzenich}}, \bibinfo {author} {\bibfnamefont {J.}~\bibnamefont {Kampmeier}}, \bibinfo {author} {\bibfnamefont {I.~E.}\ \bibnamefont {Batov}}, \bibinfo {author} {\bibfnamefont {G.}~\bibnamefont {Mussler}}, \bibinfo {author} {\bibfnamefont {J.}~\bibnamefont {Schubert}}, \bibinfo {author} {\bibfnamefont {D.}~\bibnamefont {Gr\"utzmacher}},\ and\ \bibinfo {author} {\bibfnamefont {T.}~\bibnamefont {Sch\"apers}},\ }\bibfield  {title} {\bibinfo {title} {{Magnetoresistance oscillations in {MBE-grown} {Sb$_2$Te$_3$} thin films}},\ }\href {https://doi.org/10.1063/1.4977848} {\bibfield  {journal} {\bibinfo  {journal} {Appl. Phys. Lett.}\ }\textbf {\bibinfo {volume} {110}},\ \bibinfo {pages} {092104} (\bibinfo {year} {2017})}\BibitemShut {NoStop}%
\bibitem [{\citenamefont {Yu}\ and\ \citenamefont {Cardona}(1996)}]{YuFS96}%
  \BibitemOpen
  \bibfield  {author} {\bibinfo {author} {\bibfnamefont {P.~Y.}\ \bibnamefont {Yu}}\ and\ \bibinfo {author} {\bibfnamefont {M.}~\bibnamefont {Cardona}},\ }\href@noop {} {\emph {\bibinfo {title} {Fundamentals of semiconductors: physics and materials properties}}}\ (\bibinfo  {publisher} {Springer},\ \bibinfo {year} {1996})\BibitemShut {NoStop}%
\bibitem [{\citenamefont {Eschbach}\ \emph {et~al.}(2015)\citenamefont {Eschbach}, \citenamefont {M{\l}y{\'{n}}czak}, \citenamefont {Kellner}, \citenamefont {Kampmeier}, \citenamefont {Lanius}, \citenamefont {Neumann}, \citenamefont {Weyrich}, \citenamefont {Gehlmann}, \citenamefont {Gospodari{\v{c}}}, \citenamefont {D{\"o}ring}, \citenamefont {Mussler}, \citenamefont {Demarina}, \citenamefont {Luysberg}, \citenamefont {Bihlmayer}, \citenamefont {Sch{\"a}pers}, \citenamefont {Plucinski}, \citenamefont {Bl{\"u}gel}, \citenamefont {Morgenstern}, \citenamefont {Schneider},\ and\ \citenamefont {Gr{\"u}tzmacher}}]{EschbachNC15}%
  \BibitemOpen
  \bibfield  {author} {\bibinfo {author} {\bibfnamefont {M.}~\bibnamefont {Eschbach}}, \bibinfo {author} {\bibfnamefont {E.}~\bibnamefont {M{\l}y{\'{n}}czak}}, \bibinfo {author} {\bibfnamefont {J.}~\bibnamefont {Kellner}}, \bibinfo {author} {\bibfnamefont {J.}~\bibnamefont {Kampmeier}}, \bibinfo {author} {\bibfnamefont {M.}~\bibnamefont {Lanius}}, \bibinfo {author} {\bibfnamefont {E.}~\bibnamefont {Neumann}}, \bibinfo {author} {\bibfnamefont {C.}~\bibnamefont {Weyrich}}, \bibinfo {author} {\bibfnamefont {M.}~\bibnamefont {Gehlmann}}, \bibinfo {author} {\bibfnamefont {P.}~\bibnamefont {Gospodari{\v{c}}}}, \bibinfo {author} {\bibfnamefont {S.}~\bibnamefont {D{\"o}ring}}, \bibinfo {author} {\bibfnamefont {G.}~\bibnamefont {Mussler}}, \bibinfo {author} {\bibfnamefont {N.}~\bibnamefont {Demarina}}, \bibinfo {author} {\bibfnamefont {M.}~\bibnamefont {Luysberg}}, \bibinfo {author} {\bibfnamefont {G.}~\bibnamefont {Bihlmayer}}, \bibinfo {author} {\bibfnamefont {T.}~\bibnamefont {Sch{\"a}pers}}, \bibinfo {author}
  {\bibfnamefont {L.}~\bibnamefont {Plucinski}}, \bibinfo {author} {\bibfnamefont {S.}~\bibnamefont {Bl{\"u}gel}}, \bibinfo {author} {\bibfnamefont {M.}~\bibnamefont {Morgenstern}}, \bibinfo {author} {\bibfnamefont {C.~M.}\ \bibnamefont {Schneider}},\ and\ \bibinfo {author} {\bibfnamefont {D.}~\bibnamefont {Gr{\"u}tzmacher}},\ }\bibfield  {title} {\bibinfo {title} {Realization of a vertical topological p--n junction in epitaxial {Sb$_2$Te$_3$/Bi$_2$Te$_3$} heterostructures},\ }\href {https://doi.org/10.1038/ncomms9816} {\bibfield  {journal} {\bibinfo  {journal} {Nature Comm.}\ }\textbf {\bibinfo {volume} {6}},\ \bibinfo {pages} {8816} (\bibinfo {year} {2015})}\BibitemShut {NoStop}%
\end{thebibliography}%

\end{document}